\documentclass[10pt,aps,prl,twocolumn,superscriptaddress,floatfix,showpacs]{revtex4-1}
\usepackage{graphicx}
\usepackage{amsfonts}
\usepackage{amssymb}
\usepackage{amsmath}
\usepackage{txfonts}
\usepackage{lipsum}
\usepackage{color}
\usepackage{wasysym}
\usepackage{hyperref}
\usepackage{bbold}
\usepackage{etoolbox} 

\DeclareMathOperator{\Tr}{Tr}

\newcommand{\av}[1]{\ensuremath{\left\langle #1 \right\rangle}}
\newcommand{\qv}{\mathbf{q}}
\newcommand{\kv}{\mathbf{k}}

\newcommand{\avsmall}[1]{\ensuremath{\langle #1 \rangle}}

\newcommand{\msigma}{\ensuremath{\bar{\sigma}}}
\newcommand{\dn}{\ensuremath{\downarrow}}
\newcommand{\up}{\ensuremath{\uparrow}}

\begin{document}
\title{
Effective Heisenberg model and exchange interaction for strongly correlated systems}

\author{E. A. Stepanov}
\affiliation{Radboud University, Institute for Molecules and Materials, 6525AJ Nijmegen, The Netherlands}
\affiliation{\mbox{Theoretical Physics and Applied Mathematics Department, Ural Federal University, Mira Str. 19, 620002 Ekaterinburg, Russia}}

\author{S. Brener}
\affiliation{Institute of Theoretical Physics, University of Hamburg, 20355 Hamburg, Germany}

\author{F. Krien}
\affiliation{Institute of Theoretical Physics, University of Hamburg, 20355 Hamburg, Germany}

\author{M. Harland}
\affiliation{Institute of Theoretical Physics, University of Hamburg, 20355 Hamburg, Germany}

\author{A. I. Lichtenstein}
\affiliation{Institute of Theoretical Physics, University of Hamburg, 20355 Hamburg, Germany}
\affiliation{\mbox{Theoretical Physics and Applied Mathematics Department, Ural Federal University, Mira Str. 19, 620002 Ekaterinburg, Russia}}

\author{M. I. Katsnelson}
\affiliation{Radboud University, Institute for Molecules and Materials, 6525AJ Nijmegen, The Netherlands}
\affiliation{\mbox{Theoretical Physics and Applied Mathematics Department, Ural Federal University, Mira Str. 19, 620002 Ekaterinburg, Russia}}


\begin{abstract}
We consider the extended Hubbard model and introduce a corresponding Heisenberg-like problem written in terms of spin operators. The derived formalism is reminiscent of Anderson's idea of the effective exchange interaction and takes into account nonlocal correlation effects. The results for the exchange interaction and spin susceptibility in the magnetic phase are expressed in terms of single-particle quantities. This fact can be used not only for realistic calculations of multiband systems, but also allows to reconsider a general description of many-body effects in the most interesting physical regimes where the physical properties of the system are dominated by collective (bosonic) fluctuations. In the strongly spin-polarized limit, when the local magnetic moment is well-defined, the exchange interaction reduces to a standard expression of the density functional theory that has been successfully used in practical calculations of magnetic properties of real materials. 
\end{abstract}

\maketitle
The theory of magnetism is one of the most attractive and discussed areas of physics. An additional interest to this topic is heated up by the theoretical prediction~\cite{bogdanov1989thermodynamically} and experimental observation~\cite{Muhlbauer915, yu2010real, heinze2011spontaneous} of topologically stable skyrmionic spin textures that are intensively studied nowadays in the context of spintronics and magnetic data storing~\cite{Jonietz1648, fert2013skyrmions, Romming636}. Also, a correct account for spin excitations is important for realization of Kitaev spin model~\cite{1063-7869-44-10S-S29, KITAEV20062} and its practical application in the Majorana quantum computers~\cite{doi:10.1146/annurev-conmatphys-030212-184337, PhysRevLett.105.077001, PhysRevLett.104.040502, Mourik1003, Nadj-Perge602, Reiher7555}. A quantitative description of the mentioned effects requires the knowledge of the exchange interaction between two spins. However, this problem is challenging when applied to many magnetic materials that are by definition strongly correlated quantum systems. 

Originally, the development of the theory of exchange interactions in solids and molecules was based on the Heitler-London theory of the hydrogen molecule~\cite{HL}. It has been demonstrated, however, in the early 60's by Freeman and Watson~\cite{FW} that this theory, being applied to ferromagnetic transition metals, gives a completely wrong order of magnitude and even an incorrect sign of the exchange parameters. For magnetic insulators, a semi-empirical theory of exchange interactions has been developed in the 50's, known as Goodenough-Kanamori-Anderson rules~\cite{M1,M2,M3,M4}, however, it was not quantitative. An analysis of ``superexchange'' in particular compounds always assumed some model considerations, that is, the importance and non-importance of specific intermediate states. When the density functional theory (DFT) became the base of microscopic quantum theory of molecules and crystals~\cite{HeLu,JG,Kubler} the most straightforward way to estimate the exchange interactions was simply the calculation of the total energy difference between ferromagnetic and antiferromagnetic phases. This assumes the applicability of the Heisenberg model, which is frequently not the case, especially for itinerant electron systems~\cite{Kubler,Herring,Moriya,IKS}.

A general, model-independent and parameter-free method to calculate exchange interactions within DFT was suggested in Refs.~\cite{LKG84,LKG85,LKAG87} based on the ``magnetic local force theorem''. It is based on the consideration of second-order variations of the total energy with respect to small rotations of magnetic moments starting from equilibrium  ground states. Later this approach was generalized to strongly correlated systems~\cite{KL2000, PhysRevB.94.115117} (within the framework of dynamical mean-field theory (DMFT)~\cite{PhysRevLett.62.324, RevModPhys.68.13}), magnetic systems out of equilibrium~\cite{SECCHI2013221}, and relativistic magnetic interactions, such as the Dzyaloshinskii-Moriya interaction~\cite{KKML10,SLK15,SLK16a}. This theory was successfully used for many calculations of real systems, such as magnetic semiconductors~\cite{MagSem}, molecular magnets~\cite{MolMag1,MolMag2}, ferromagnetic transition metals~\cite{Fe1,Fe2} and half-metallic ferromagnets~\cite{CrO2}.

Despite the success of this approach its conceptual status remains unclear. Indeed, a mapping from DFT or from a  Hubbard model to the Heisenberg model is in general impossible; exchange interactions obtained from the magnetic force theorem are classical and dependent on the magnetic configuration (see, e.g.~\cite{remko}). Their relation to observables is not very clear; strictly speaking, only the spin-wave stiffness constant in ferromagnets is a well-defined quantity since we can be sure that in the limit of slow times and large spatial scales the phenomenological Landau-Lifshitz equations are correct. This was emphasized already in the first paper~\cite{LKG84}. 
Observables are directly related to the dynamic magnetic susceptibility, but to establish relations between the magnetic local force approach and the standard language of response functions is not an easy problem. It was solved only within the local spin-density approximation in DFT~\cite{KL04} and within the time-dependent mean-field approach in the Hubbard model~\cite{SLK16b}. However, most of the interesting magnetic materials are strongly correlated systems, and these approximations seem to be insufficient (or, at least, not completely justified) to describe spin dynamics.

In this Letter we show that the extended Hubbard Hamiltonian can be mapped onto an effective Heisenberg model. Inspired by the Dual Boson (DB) formalism~\cite{Rubtsov20121320, PhysRevB.90.235135, PhysRevB.93.045107, PhysRevB.94.205110} we construct a bosonic model, whose interaction is reminiscent of Anderson's superexchange mechanism~\cite{PhysRev.115.2,PhysRevB.52.10239}. Importantly, the derived formalism remains applicable not only in the strongly localized regime and allows the description of every magnetic system with a well-defined local magnetic moment. Moreover, the presence of the latter
allows to reveal a general way of the description of a complicated quantum many-body problem in terms of single-particle quantities with the use of Ward identities~\cite{Hafermann14-2,KrienF}.

{\it Effective $s$-$d$ model} --- 
We consider the action of the extended Hubbard model for correlated electrons,
\begin{align}
\label{eq:actionlatt}
{\cal S} = &-\sum_{\kv,\nu,\sigma} c^{*}_{\kv\nu\sigma} \left[i\nu+\mu-\varepsilon^{\phantom{\dagger}}_{\kv}\right]
\, c^{\phantom{*}}_{\kv\nu\sigma} \\
&+ U\sum\limits_{\qv,\omega} n^{*}_{\qv\omega\uparrow} n^{\phantom{*}}_{\qv\omega\downarrow} +\frac12\sum\limits_{\qv,\omega,\varsigma}\rho^{*\,\varsigma}_{\qv\omega} \left[V^{\phantom{*}}_{\qv}\right]_{\varsigma\varsigma}\rho^{\,\varsigma}_{\qv\omega}. \notag
\end{align}
Here $c^{*}_{\kv\nu\sigma}$ ($c_{\kv\nu\sigma}$) are Grassmann variables corresponding to creation (annihilation) of an electron with momentum $\kv$,  fermionic Matsubara frequency $\nu$ and spin $\sigma$ labels.
The label $\varsigma=\{c, {\bf s}\}$ depicts charge $c$ and spin ${\bf s} = \{x,y,z\}$ degrees of freedom, so that $U$ corresponds to local Coulomb interaction, $[V_{\qv}]_{cc}=V_{\qv}$ and $[V_{\qv}]_{ss}=-J^{\rm d}_{\qv}/2$ describe nonlocal Coulomb and direct ferromagnetic exchange interactions, respectively. Here, we also introduce bosonic variables:  $\rho^{\,\varsigma}_{\qv\omega}=n^{\,\varsigma}_{\qv\omega}-\langle n^{\,\varsigma}_{\qv\omega}\rangle$, where $n^{\varsigma}_{\qv\omega} = \sum_{\kv\nu\sigma\sigma'}c^{*}_{\kv\nu\sigma} \sigma^{\varsigma}_{\sigma\sigma'} c^{\phantom{*}}_{\kv+\qv,\nu+\omega,\sigma'}$ is the charge ($\varsigma=c$) and spin ($\varsigma=s$) density of electrons with the momentum $\qv$, bosonic frequency $\omega$ and Pauli matrices $\sigma^{\varsigma}=\{\mathbb{1},\boldsymbol{\sigma}^{s}\}$.  

Expressing the effective exchange interaction in terms of correlation functions is a nontrivial task, since it is not an observable. Furthermore, in the strongly correlated regime charge and spin fluctuations are entangled in a complicated way.
Both challenges can be approached within the Dual Boson formalism~\cite{Rubtsov20121320, PhysRevB.90.235135, PhysRevB.93.045107, PhysRevB.94.205110}, since it naturally separates charge and spin degrees of freedom by representing them in terms of bosonic fields entering an effective action. To this aim one splits the lattice action~\eqref{eq:actionlatt} into the local impurity problem 
of the extended dynamical mean-field theory (EDMFT,~\cite{PhysRevB.52.10295, PhysRevLett.77.3391, PhysRevB.61.5184, PhysRevLett.84.3678, PhysRevB.63.115110, PhysRevLett.90.086402}) and the remaining non-local part, 
which is a bilinear function of $c^{*}(c)$ and $\rho$ variables. 
Within the DB approach this remaining part is decoupled by two Hubbard-Stratonovich transformations, thus introducing {\it dual} fermionic $f^{*}\,(f)$ and bosonic $\phi$ fields. Then, the initial fermionic degrees of freedom $c^{*}\,(c)$ can be integrated out, leading to the interaction part $\tilde{W}[f,\phi]$ of the resulting {\it dual} action being expressed in terms of the full vertex functions of the local impurity problem (for details see Suppl. Mat.~\cite{SM}). Thus, by construction, local correlations are already embedded into the bare propagators and interactions of the DB problem, which is very convenient for practical calculations. In the following we restrict ourselves to the lowest order terms in $\tilde{W}[f,\phi]$ stemming from the four-point $\overline{\gamma}_{\nu\nu'\omega}$ and three-point $\gamma_{\nu\omega}$ vertices~\cite{SM}.

Dual fields $f^{*}\,(f)$ and $\phi$ have no direct physical interpretation, but this fact does not represent a significant obstacle for the calculation of physical observables, since there is an exact connection between dual and lattice quantities~\cite{Rubtsov20121320, PhysRevB.90.235135, PhysRevB.93.045107, PhysRevB.94.205110}. However, for our goal of deriving an effective bosonic model that describes initial (lattice) degrees of freedom it is crucial to formulate the problem in terms of bosonic fields that have a clear physical meaning. To remedy this problem, we perform the reverse Hubbard-Stratonovich transformation for the bosonic variables $\phi$ introducing fields $\bar{\rho}$. In this we were inspired by works of Dupuis~\cite{doi:10.1142/S0217979200002430, DUPUIS2001617, 2001cond.mat.DUPUIS}, where a similar trick was performed for fermionic degrees of freedom. After integrating over {\it dual} bosonic fields $\phi$ one gets the following action reminiscent of the $s$-$d$ model~\cite{SM}
\begin{align}
{\cal S}_{s\text{-}d} = &-\sum_{\kv,\nu,\sigma} f^{*}_{\kv\nu\sigma}\tilde{G}^{-1}_{0}f^{\phantom{*}}_{\kv\nu\sigma}
-\frac12\hspace{-0.1cm}\sum_{\qv,\omega,\varsigma(')} \bar\rho^{*\,\varsigma}_{\qv\omega}\left[X^{\phantom{*}}_{\rm E}\right]^{-1}_{\varsigma\varsigma'}\bar\rho^{\,\varsigma'}_{\qv\omega} + W.
\label{eq:actionSf}
\end{align}
Here, $X_{\rm E}$ is the EDMFT susceptibility and $\tilde{G}_0$ is the nonlocal part of the EDMFT Green's function. Importantly, after all transformations the field $\bar{\rho}$ indeed has the same physical meaning as original {\it composite} bosonic field $\rho$ of the lattice problem~\eqref{eq:actionlatt} as shown in~\cite{SM}. The decisive advantage of the variable $\bar{\rho}$ is that it can now be treated as the {\it elementary} bosonic field that has a well-defined propagator and is independent of fermionic degrees of freedom $c^{*}\,(c)$. Remarkably, $W[f,\bar\rho]$ keeps the practical form of the dual interaction $\tilde{W}[f,\phi]$ with the replacement of bosonic variable $\phi\to\bar\rho$, although the four-fermionic term is modified under these transformations. As we argue in~\cite{SM} and numerically check below, in the case of well-developed bosonic fluctuations this modification results in the corresponding contribution to the interaction $W[f,\rho]$ becoming negligibly small and the latter takes the simple form 
$W[f,\rho]
\simeq\sum_{\kv,\qv}\sum_{\nu,\omega,\varsigma}\gamma^{\,\varsigma}_{\nu\omega}\,\rho^{*\,\varsigma}_{\qv\omega} 
 f^{*}_{\kv\nu\sigma} f^{\phantom{*}}_{\kv+\qv,\nu+\omega,\sigma'}$. 
At last we mention, that the fermionic degrees of freedom are kept in the dual space, which will prove to be useful to discriminate between local and nonlocal contributions to the lattice susceptibility.

{\it Magnetic susceptibility} ---
In order to design an effective Heisenberg model for spin degrees of freedom, one has to assume that the local magnetization $\av{m}=2\av{S^{z}}$ is described well at the dynamical mean-field level and fluctuations revealed by the system beyond EDMFT are mostly bosonic. In order to have well-defined local magnetic moment, the effective impurity model has to be considered for the spin polarized state. For easier description, one can transform spin variables from ${\bf s}=\{x,y,z\}$ to ${\bf s}=\{+,-,z\}$ basis with $S^{\pm}=(\rho^{x}\pm{}i\rho^{y})/2$. In the spin-polarized case charge and spin $z$ channels are yet entangled, but the $\pm$ spin channel can be separated in the collinear case~\cite{BICKERS1989206, 0953-8984-11-4-011}. Thus, for the correct description of the spin fluctuations, one may consider correlations only in the $\pm$ spin channel and the contribution of the $z$ channel to the exchange interaction can be later restored from symmetry arguments. For simplicity, $\pm$ spin labels are omitted wherever they are not crucial for understanding.  

Now, one can integrate out fermionic degrees of freedom in the effective action~\eqref{eq:actionSf} and get the following spin model
\begin{align}
{\cal S}_{\rm spin}
&=-\frac12\sum_{\qv,\omega} S^{\,-}_{\qv\omega} 
\left[X_{\qv\omega}^{\,-+}\right]^{-1} \, S^{\,+}_{-\qv,-\omega} + {\rm h.c.}
\label{eq:Spm}
\end{align}
A first approximation for the magnetic susceptibility $X_{\qv\omega}$ can be obtained for the case when the interaction $W[f,\bar\rho]$ contains only the three-point vertex $\gamma^{\pm}_{\nu\omega}$, as discussed above. Therefore,
the expansion of the partition function of the action~\eqref{eq:actionSf} up to the second order with respect to bosonic fields gives~\cite{SM}
\begin{align}
\left[X^{(2)}_{\qv\omega}\right]^{-1}
&= J^{\rm d}_{\bf q} + \Lambda^{\phantom{1}}_{\omega} + \chi^{-1}_{\omega} - \tilde{\Pi}^{(2)}_{\qv\omega}.
\label{eq:X1}
\end{align}
Here, $\Lambda_{\omega}$ and $\chi_{\omega}$ are the bosonic hybridization function and susceptibility of the impurity problem, respectively. Also,
\begin{align}
\tilde{\Pi}^{(2)}_{\qv\omega}=\sum_{\kv,\nu}
\gamma^{\,-}_{\nu+\omega,-\omega}\,\tilde{G}^{\phantom{2}}_{\kv+\qv,\nu+\omega\uparrow}\tilde{G}^{\phantom{2}}_{\kv\nu\downarrow}\,\gamma^{\,+}_{\nu,\omega}=
\vcenter{\hbox{\includegraphics[width=0.15\linewidth]{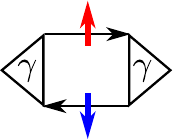}}}
\label{eq:Pi2fig}
\end{align}
is the second order polarization function~\cite{PhysRevB.90.235135}. 
Note that a conserving description of spin fluctuations is given by the two-particle ladder approximation of the magnetic susceptibility provided by the ladder DB approach~\cite{PhysRevB.93.045107} that accounts for the four-fermionic contribution in $W[f,\bar\rho]$ and treats bosonic hybridization $\Lambda$ as a constant~\cite{KrienF}
\begin{align}
\left[X^{\rm ladd}_{\qv\omega}\right]^{-1} = J^{\rm d}_{\qv} + \Lambda + \left[X^{\rm DMFT}_{\qv\omega}\right]^{-1}.
\label{eq:X_DB}
\end{align}
Here, $X^{\rm DMFT}_{\qv\omega}=\chi^{\phantom{2}}_{\omega}+\chi^{\phantom{2}}_{\omega}\tilde{\Pi}^{\rm ladd}_{\qv\omega}\chi^{\phantom{2}}_{\omega}$ is the DMFT-~\cite{PhysRevLett.62.324, RevModPhys.68.13}, or D$\Gamma$A-like~\cite{PhysRevB.75.045118} susceptibility written in terms of local two-particle irreducible four-point vertices and lattice Green's functions. $\tilde{\Pi}^{\rm ladd}_{\qv\omega}$ is the dual polarization in the ladder form~\cite{PhysRevB.90.235105,SM} that contains $\tilde{\Pi}^{(2)}_{\qv\omega}$ as the lowest order term. 
Therefore, the hybridization $\Lambda$ plays the role of the Moriyaesque $\lambda$ correction that was introduced in D$\Gamma$A~\cite{PhysRevB.80.075104} by hand similarly to the Moriya and Kawabata theory of weak itinerant magnets~\cite{1973639, 1973669} and now is derived analytically.

\begin{figure}[t!]
\includegraphics[width=0.9\linewidth]{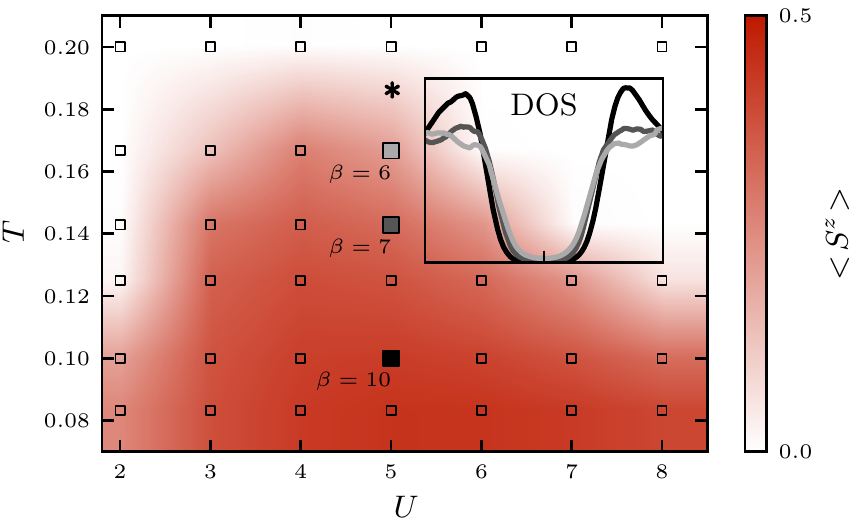}
\vspace{-0.3cm}
\caption{\label{fig:pd} (Color online) The antiferromagnetic phase of the half-filled Hubbard model. Squares mark where calculations were done, the red color depicts the magnitude of the magnetic moment $\av{S^z}$ and the asterisk marks the N\'eel temperature $T_N\approx0.186$. The inset shows the total DOS at $\beta=6, 7$ and $10$ for $U=5$.
}
\end{figure}

Importantly, the expressions for the magnetic susceptibility~\eqref{eq:X1} and~\eqref{eq:X_DB} can be drastically simplified to be applicable for realistic multiband calculations, for which the two-particle quantities can hardly be obtained. As it was discussed above, the system with a well-defined local magnetic moment exhibits mostly bosonic fluctuations. Therefore, one can expect that local vertex functions are mostly described by the bosonic frequency $\omega$, while the dependence on fermionic frequencies $\nu,\nu'$ is negligible and can be averaged out. In order to perform this averaging consistently, it is carried out using the local Ward identities~\cite{Hafermann14-2,KrienF}, which leads to the following approximation of three-point vertex~\cite{SM}
\begin{align}
\gamma^{+}_{\nu\omega} = \gamma^{-}_{\nu+\omega,-\omega} \simeq \chi^{-1}_{\omega} + \delta\Sigma^{\phantom{1}}_{\nu\omega}~\simeq \chi^{0~-1}_{\omega}
\label{eq:3vertex}
\end{align}
Here, $\chi^{0}_{\omega}=\sum_{\nu}g_{\nu+\omega\uparrow}g_{\nu\downarrow}$ is the bare spin susceptibility, $g_{\nu\sigma}$ and $\Sigma_{\nu\sigma}$ are the full Green's function and self-energy of the impurity problem and $\delta\Sigma_{\nu\omega} = (\Sigma_{\nu+\omega\uparrow}-\Sigma_{\nu\downarrow})/\av{m}$. Therefore, exploiting the system being in the magnetic phase allows to rewrite the complicated many-body problem~\eqref{eq:actionlatt} in a much simpler form of Eq.~\ref{eq:actionSf} introducing bosonic fields that correspond to the collective magnetic fluctuations. In this case, the expression for the corresponding fermion-boson coupling $\gamma^{\pm}_{\nu\omega}$ can be in drastically simplified~\eqref{eq:3vertex}, leading to a similar expression that was recently postulated in~\cite{Scheurer201720580} and numerically checked using brute force calculations~\cite{2017arXiv170706602W}.

{\it Exact numerical solution} --- 
In order to exemplify the above approximations we consider the half-filled Hubbard model~\eqref{eq:actionlatt} ($V_{\qv}, J^{\rm d}_{\qv}, \Lambda=0$) on the hypercubic lattice in infinite dimensions. In this case, the exact result for the magnetic susceptibility is known to be given by the DMFT expression~\eqref{eq:X_DB} and can be compared to the simplified result of Eq.~\ref{eq:X1}. At low temperatures this system favors antiferromagnetic (AFM) order over paramagnetism as shown in the phase diagram in Fig.~\ref{fig:pd}.

\begin{figure}[t]
\includegraphics[width=0.95\linewidth]{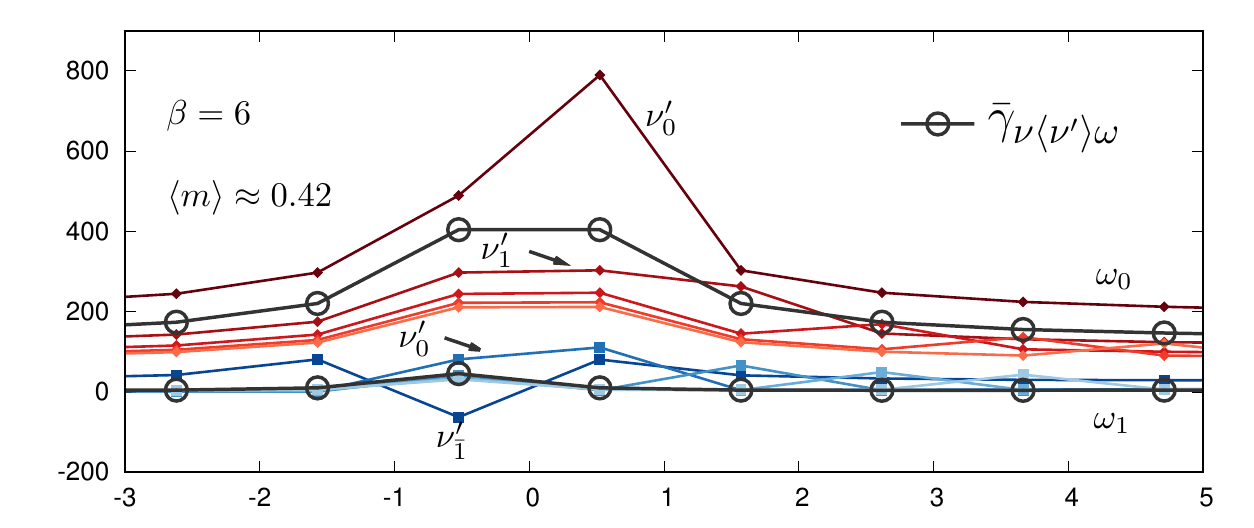}
\includegraphics[width=0.95\linewidth]{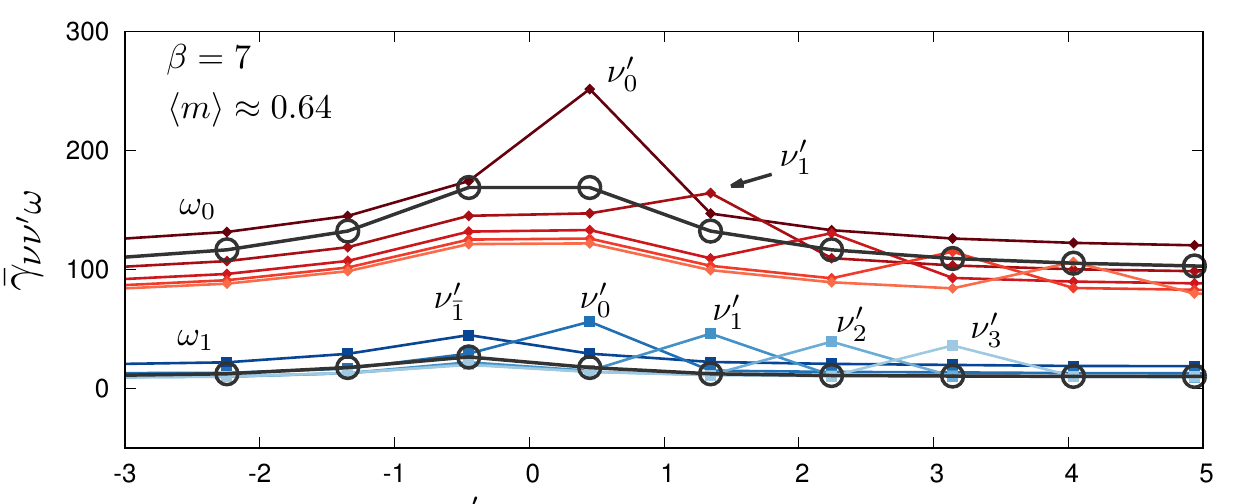}
\includegraphics[width=0.95\linewidth]{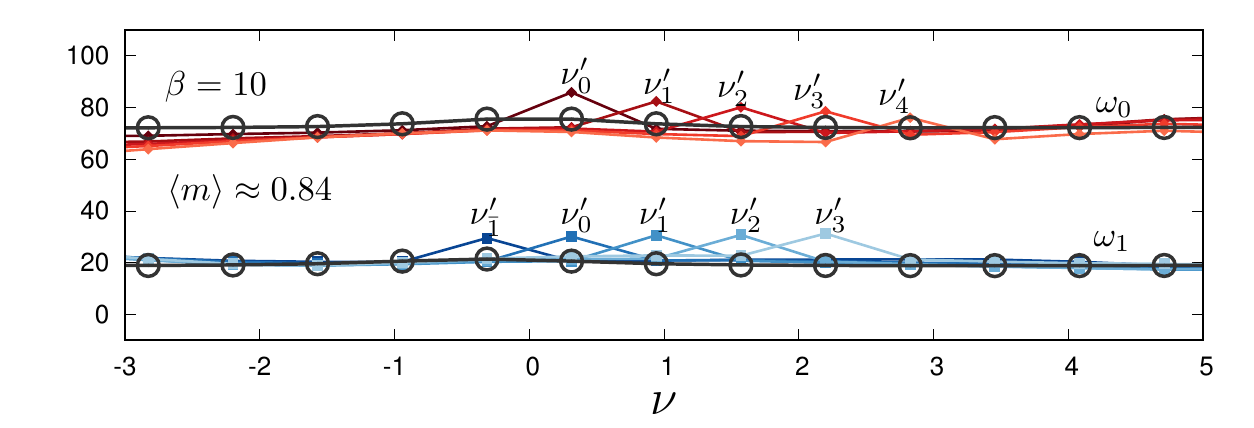}
\vspace{-0.3cm}
\caption{\label{fig:gamma_av} (Color online) Real part of the four-point vertex $\overline{\gamma}_{\nu\nu'\omega}$ in the $\pm$ spin channel at $U=5$ for three different temperatures (cf. marked points in Fig.~\ref{fig:pd}). The plot shows $\overline{\gamma}_{\nu\nu'\omega}$ as a function of $\nu$ for fixed $\omega$ and $\nu'$. Diamonds and squares show data for $\omega=\omega_0$ and $\omega_1$, respectively. Red ($\omega_0$) and blue ($\omega_1$) lines serve as guides to the eye, lighter colors indicate larger $\nu'$. Black circles and lines show $\overline{\gamma}_{\nu\av{\nu'}\omega}$, which does not depend on $\nu'$. 
}
\end{figure} 

The local four-point vertex $\overline{\gamma}_{\nu\nu'\omega}$ is measured at $U=5$ for the three temperatures marked in Fig.~\ref{fig:pd}, roughly below the maximum of the AFM dome, where $T_N\approx0.186$ is obtained using DMFT~\cite{SM}. As the temperature is lowered from $\beta=6$ to $10$, the magnetization increases from $\av{m}\simeq0.42$ to $0.84$. We validate in Fig.~\ref{fig:gamma_av} that at large magnetization the dependence of the four-point vertex $\overline{\gamma}_{\nu\nu'\omega}$ on fermionic frequencies $\nu,\nu'$ is small. Consequently, one may indeed use the approximated form of the vertex $\overline\gamma_{\nu\nu'\omega}\simeq\overline\gamma_{\nu\av{\nu'}\omega}$, which leads to Eq.~\ref{eq:3vertex}.

\begin{figure}[t]
\includegraphics[width=0.9\linewidth]{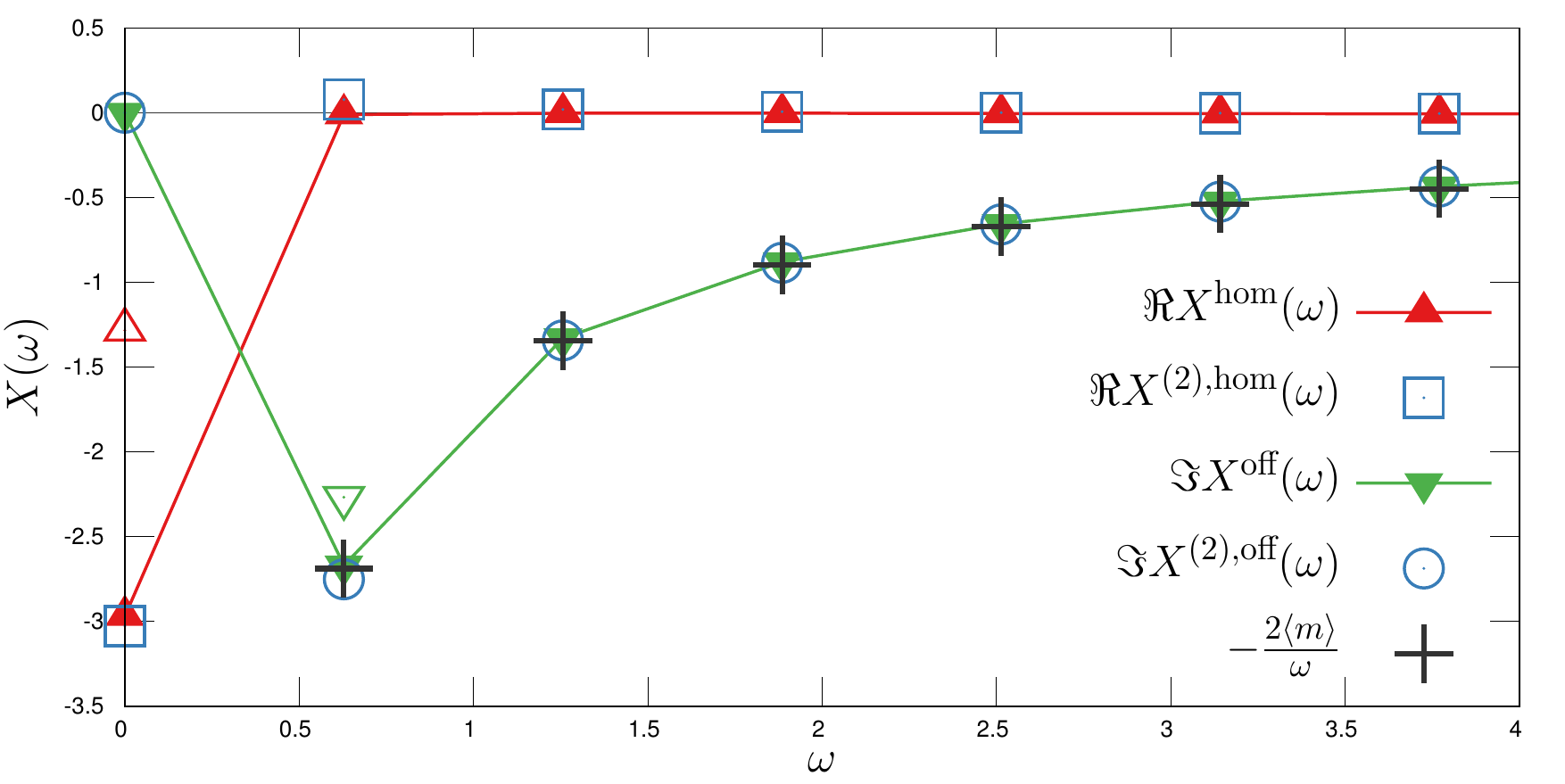}
\vspace{-0.3cm}
\caption{\label{fig:susc} (Color online) Spin susceptibility components $X^{\text{hom}}_{\omega}$ and $X^{\text{off}}_{\omega}$ as a function of the Matsubara frequency (triangles). Squares and circles show the simplified form of the magnetic susceptibility~\eqref{eq:X1}. The single red triangles indicate expression for the magnetic susceptibility in the case of the truncated ladder (see text). The parameters of this figure correspond to the bottom panel of Fig.~\ref{fig:gamma_av}.} 
\end{figure}

We evaluate Eq.~\eqref{eq:X_DB} in the AFM phase at the $\qv=0$ point of the reduced Brillouin zone. The transversal susceptibility is a $2\times2$ matrix with the homogeneous susceptibility $X^{\text{hom}}(\omega)$ as a diagonal element~\cite{Note1}. Fig.~\ref{fig:susc} shows $X^{\text{hom}}(\omega)$, which is real, as well as the off-diagonal element $X^{\rm off}(\omega)$. Remarkably, despite the approximation of the vertex functions, $X^{\rm hom}(\omega\neq0)=0$ and $X^{\rm off}(\omega\neq0)=-2i\av{m}/\omega$ hold to very good accuracy, which are exact constraints due to global spin conservation~\cite{SM}. 

At $U=5$ the eigenvalue of the ladder Eq.~\eqref{eq:X_DB} corresponding to $X^{\text{hom}}(\omega=0)$
is large ($\simeq0.715$). Therefore, one can not approximate the polarization $\tilde{\Pi}^{\rm ladd}_{\qv\omega}$ by the second order expression $\tilde\Pi^{(2)}_{\qv\omega}$ in Eq.~\ref{eq:X_DB}. The corresponding approximation for $\Re X^{\text{hom}}(\omega=0)$ and $\Im X^{\text{off}}(\omega=2\pi\beta)$ is marked in Fig.~\ref{fig:susc} with open triangles and indeed clearly distinguishable from Eq.~\eqref{eq:X_DB}.

Nonetheless, the simplified expression for magnetic susceptibility $X^{(2)}$~\eqref{eq:X1} with the vertex approximation~\eqref{eq:3vertex} shows a good agreement with $X^{\rm ladd}$~\eqref{eq:X_DB}.
Importantly, the approximation for the magnetic susceptibility obtained in Eq.~\ref{eq:X1} should not be confused with the truncation of the ladder equation, even though it formally uses the same quantity $\tilde{\Pi}^{(2)}_{\qv\omega}$. The good agreement of the simplified result $X^{(2)}$ with the much more advanced ladder approximation~\eqref{eq:X_DB} shows that the bosonic fluctuations indeed dominate in the polarized regime of the impurity model, which was assumed while deriving Eq.~\ref{eq:X1}. 

{\it Classical Heisenberg Hamiltonian} --- Although the action~\eqref{eq:Spm} is general and can be used for the description of quantum effects in terms of susceptibilities, at low temperatures it can be mapped onto an effective classical Heisenberg Hamiltonian $H_{\rm spin} = - \sum_{\qv} J_{\qv}\,{\bf S}_{\qv}\,{\bf S}_{-\qv}$ that describes small spin fluctuations around the AFM ground state~\cite{LKG85}. To this aim, spin variables $S^{\pm}_{\qv\omega}$ in~\eqref{eq:Spm} are replaced by classical vectors ${\bf S}_{\qv}$ of the length $\langle S_{z} \rangle$ and the contribution from the $z$ spin channel is restored from the requirement of rotational invariance. Then, an effective exchange interaction $J_{\qv}$ can be defined as a nonlocal part of the inverse spin susceptibility at the zero bosonic frequency~\cite{KL2000}. 
Thus, the effective exchange interaction that corresponds to the simplified form of magnetic susceptibility~\eqref{eq:X1} reads
\begin{align}
J^{\phantom{d}}_{\qv} = J^{\rm d}_{\qv} - \sum_{\kv,\nu}
\gamma^{\,-}_{\nu,\omega=0}\,\tilde{G}^{\phantom{2}}_{\kv+\qv,\nu\uparrow}\tilde{G}^{\phantom{2}}_{\kv\nu\downarrow}\,\gamma^{\,+}_{\nu,\omega=0},
\label{eq:Jfinal}
\end{align}
while the exchange interaction in the ladder approximation is detailed in~\cite{SM}. 
This result reminds of Anderson's idea of the superexchange interaction~\cite{PhysRev.115.2,PhysRevB.52.10239}. Indeed, the first and the second term in Eq.~\ref{eq:Jfinal} describe the direct ferromagnetic and kinetic antiferromagnetic exchange interactions, respectively. As a result, in the strongly localized regime and in the case of antiferromagnetic dimer the kinetic part of the exchange interaction takes the well-known form $J=-2t^2/U$~\cite{SM}.

It is worth mentioning that the three-point vertex $\gamma_{\nu,\omega}$ that enters the kinetic part of the exchange interaction describes the total spin splitting. In the spin polarized case one can again use the simplified form of the vertex function (first approximation in Eq.~\ref{eq:3vertex}). In the strongly polarized regime the potential contribution to the spin splitting $\delta\Sigma_{\nu\omega}$ is much larger than the kinetic one $\chi^{-1}_{\omega}$. Therefore, the latter can be neglected and the result for the exchange interaction~\eqref{eq:Jfinal} reduces to the expression obtained in~\cite{KL2000} that was successfully applied to the description of many realistic systems~\cite{MagSem,MolMag1,MolMag2,Fe1,Fe2,CrO2}. Note that in~\cite{KL2000} the exchange interaction was derived assuming the existence of the collinear spin ground state, while here we show that the limit of applicability of the derived expression is much broader. If the dependence of the three-point vertex on the fermionic frequencies is fully disregarded (second approximation in Eq.~\ref{eq:3vertex}), the exchange interaction reduces to the ``Hartree-Fock'' approximation $J_{\qv} = \chi^{0~-1}_{\omega=0}\,X^{\,0}_{\qv,\omega=0}\,\chi^{0~-1}_{\omega=0}$~\cite{SM} derived in~\cite{ANTROPOV2003L192}.

{\it Conclusion} --- 
To conclude, here we derived the action for effective $s$-$d$ and Heisenberg-like problems for the extended Hubbard model. We observed that by virtue of a local Ward identity the vertex functions of the impurity model can be well approximated, provided its weak dependence on the fermionic frequencies. Our results show that this criterion is indeed satisfied in the AFM phase of the Hubbard model in infinite dimensions when the staggered magnetization is sufficiently large. As a consequence, it is possible to obtain the magnetic susceptibility without a costly measurement of the impurity vertex functions, which is very useful for the realistic multiband calculations. 
For the considered parameters this approximation becomes accurate enough to reach an agreement with the global spin conservation. In finite dimensions this is of importance for a sound description of magnon spectra in accord with Goldstone's theorem.
In the classical limit, the derived spin action reduces to an effective Heisenberg Hamiltonian. In the spin-polarized case the result for the kinetic part of the effective exchange interaction simplifies to the expression derived in~\cite{KL2000}, which is argued to be a good approximation for the case of many real materials. 
We believe that this approximation can be applied in different and, in particular, more realistic contexts. We further speculate that similar approximations could prove valuable in any physical regime where it can be argued that the behavior of the vertex functions is strongly dominated by the transferred momentum.

\begin{acknowledgments}
The authors thank Hugo Strand and Vladimir Antropov for inspiring discussions. We kindly thank Igor Krivenko for providing the solver for stochastic optimization method for analytical continuation~\cite{som, PhysRevB.62.6317, 2018arXiv180503521G} used to obtain the data shown in Fig.~\ref{fig:pd}. We also would like to thank Junya Otsuki for providing the CTQMC solver~\cite{Otsuki13} that was used to obtain the data shown in Fig.~\ref{fig:gamma_av}. The work of E.A.S. was supported by the Russian Science Foundation, Grant 18-12-00185. The work of M.I.K. was supported by NWO via Spinoza Prize and by ERC Advanced Grant 338957 FEMTO/NANO. M.H., F.K., and A.I.L. acknowledge support from the excellence  cluster ``The Hamburg Centre for Ultrafast Imaging - Structure, Dynamics and Control of Matter at the Atomic Scale'' and North-German Supercomputing Alliance (HLRN) under the Project No. hhp00040. Also, the work was partially supported by the Stichting voor Fundamenteel Onderzoek der Materie (FOM), which is financially supported by the Nederlandse Organisatie voor Wetenschappelijk Onderzoek (NWO).
\end{acknowledgments}

\bibliographystyle{apsrev4-1}
\bibliography{DBExchange.bib}

\appendix
\clearpage
\onecolumngrid

\begin{center}
\Large{Supplemental Material for \\
``Effective Heisenberg model and exchange interaction for strongly correlated systems''}
\end{center}

\section*{Effective spin problem for the extended Hubbard model}

Here we explicitly derive a spin problem for the extended Hubbard model and obtain magnetic susceptibility. For this reason, let us consider the following action written in momentum space
\begin{align}
{\cal S} = &-\sum_{\kv,\nu,\sigma} c^{*}_{\kv\nu\sigma}\left[i\nu+\mu-\varepsilon^{\phantom{*}}_{\kv}\right]
\,c^{\phantom{\dagger}}_{\kv\nu\sigma}
+ U\sum\limits_{\qv,\omega}n^{\phantom{*}}_{\qv\omega\uparrow}n^{\phantom{*}}_{-\qv,-\omega\downarrow}
+\frac12\sum\limits_{\qv,\omega,\varsigma}\rho^{*\,\varsigma}_{\qv\omega} \left[V^{\phantom{*}}_{\qv}\right]_{\varsigma\varsigma}\rho^{\,\varsigma}_{\qv\omega}. \label{eq:actionlattapp}
\end{align}
Here, $c^{*}_{\kv\nu\sigma}$ ($c_{\kv\nu\sigma}$) are Grassmann variables corresponding to creation (annihilation) of an electron with momentum $\kv$, fermionic Matsubara frequency $\nu$ and spin $\sigma$. Quantities $\varepsilon_{\kv}$ and $[V_{\qv}]_{\varsigma\varsigma}$ are the Fourier transforms of the hopping amplitude and nonlocal part of an interaction written in the matrix form, respectively. The label $\varsigma=\{c, {\bf s}\}$ depicts the charge $c$ and spin ${\bf s}=\{x,y,z\}$ degrees of freedom, so that $U$ and $[V_{\qv}]_{cc}=V_{\qv}$ describe the local and nonlocal parts of the Coulomb interaction respectively, and $[V_{\qv}]_{ss}=-J^{\rm d}_{\qv}/2$ is the nonlocal direct ferromagnetic exchange interaction. The latter ensures the following form of the Heisenberg Hamiltonian $H=-\sum_{\qv}J^{\rm d}_{\qv} \,S^{\phantom{d}}_{\qv}S^{\phantom{d}}_{-\qv}$. Here, we also introduce bosonic variables  $\rho^{\,\varsigma}_{\qv\omega}=n^{\,\varsigma}_{\qv\omega}-\av{n^{\,\varsigma}_{\qv\omega}}$, where $n^{\varsigma}_{\qv\omega} =~\sum_{\kv\nu\sigma\sigma'}c^{*}_{\kv\nu\sigma} \sigma^{\varsigma}_{\sigma\sigma'} c^{\phantom{*}}_{\kv+\qv,\nu+\omega,\sigma'}$ is the charge ($\varsigma=c$) and spin ($\varsigma=s$) density of electrons with the momentum $\qv$, bosonic frequency $\omega$ and Pauli matrices $\sigma^{\varsigma}=\{\mathbb{1},\boldsymbol{\sigma}^{s}\}$.  

Following the standard procedure of the Dual Boson theory~\cite {Rubtsov20121320, PhysRevB.90.235135, PhysRevB.93.045107, PhysRevB.94.205110}, the lattice action is divided into the local impurity ${\cal S}_{\rm imp}$ and nonlocal ${\cal S}_{\rm rem}$ parts as
\begin{align}
{\cal S}_{\rm imp} &= 
-\sum\limits_{\nu,\sigma}c^{*}_{\nu\sigma}\left[i\nu+\mu-\Delta^{\phantom{*}}_{\nu}\right]c^{\phantom{*}}_{\nu\sigma} 
+U\sum_{\omega}n^{\phantom{*}}_{\omega\uparrow} n^{\phantom{*}}_{-\omega\downarrow} 
+\frac12\sum_{\omega,\varsigma}\rho^{*\,\varsigma}_{\omega}\left[\Lambda^{\phantom{*}}_{\omega}\right]_{\varsigma\varsigma}\rho^{\,\varsigma}_{\omega}, \label{eq:actionimpapp}\\
{\cal S}_{\rm rem} &= 
-\sum\limits_{\bf{k},\nu,\sigma}c^{*}_{\bf{k}\nu\sigma}
\left[\Delta^{\phantom{*}}_{\nu}-\varepsilon^{\phantom{*}}_{\kv}\right]
c^{\phantom{*}}_{\bf{k}\nu\sigma} +
\frac12\sum_{\qv,\omega,\varsigma}\rho^{*\,\varsigma}_{\qv\omega}
\left[V^{\phantom{*}}_{\qv}-\Lambda^{\phantom{*}}_{\omega}\right]_{\varsigma\varsigma}\rho^{\,\varsigma}_{\qv\omega} + \sum\limits_{\qv,\omega,\varsigma} 
j^{\,*\,\varsigma}_{\qv\omega}\,\rho^{\,\varsigma}_{\qv\omega},
\label{eq:Simp_SM}
\end{align}
where we introduced fermionic $\Delta_{\nu}$ and bosonic $[\Lambda_{\omega}]_{\varsigma\varsigma}$ hybridization functions and sources $j^{\,\varsigma}_{\qv\omega}$ for bosonic variables.
Since here we consider a spin-polarized case of local impurity model, the fermionic hybridization function $\Delta_{\nu\sigma}$ becomes spin-dependent. The partition function of our problem is given by the following relation
\begin{align}
{\cal Z}=\int D[c^{*},c] \, e^{-{\cal S}},
\end{align}
where ${\cal S}$ is the lattice action introduced in Eq.~\ref{eq:actionlattapp}.
Using a matrix form of the Hubbard--Stratonovich transformation of the remainder term ${\cal S}_{\rm rem}$~\eqref{eq:Simp_SM} one can introduce ${\it dual}$ fermionic $f^{*},f$ and bosonic variables $\phi^{\,\varsigma}$ 
\begin{align}
\exp\left\{\sum_{\kv,\nu,\sigma} c^{*}_{\kv\nu\sigma}[\Delta^{\phantom{*}}_{\nu\sigma}-\varepsilon^{\phantom{*}}_{\kv}]c^{\phantom{*}}_{\kv\nu\sigma}\right\} &= D_{f}
\int D[f^{*},f]\exp\left\{-\sum_{\kv,\nu,\sigma}\left( f^{*}_{\kv\nu\sigma}[\Delta^{\phantom{*}}_{\nu\sigma}-\varepsilon^{\phantom{*}}_{\kv}]^{-1}f^{\phantom{*}}_{\kv\nu\sigma} + c^{*}_{\kv\nu\sigma}f^{\phantom{*}}_{\kv\nu\sigma} + f^{*}_{\kv\nu\sigma}c^{\phantom{*}}_{\kv\nu\sigma}\right)\right\},\\
\exp\left\{\sum_{\qv,\omega,\varsigma(')}\frac12\,\rho^{*\,\varsigma}_{\qv\omega}\left[\Lambda^{\phantom{*}}_{\omega}-V^{\phantom{*}}_{\qv}\right]_{\varsigma\varsigma'}\rho^{\,\varsigma'}_{\qv\omega}\right\} &= D_{\phi}
\int D[\phi]\exp\left\{-\sum_{\qv,\omega,\varsigma(')}\left( \frac12\,\phi^{*\,\varsigma}_{\qv\omega}\left[\Lambda^{\phantom{*}}_{\omega}-V^{\phantom{*}}_{\qv}\right]_{\varsigma\varsigma'}^{-1}\phi^{\,\varsigma'}_{\qv\omega} + 
\phi^{*\,\varsigma}_{\qv\omega}\,\rho^{\,\varsigma}_{\qv\omega}\right)\right\},
\end{align}
where terms $D_{f} = {\rm det}(\Delta_{\nu\sigma}-\varepsilon_{\kv})$ and $D^{-1}_{\phi} = \sqrt{{\rm det}[\Lambda_{\omega}-V_{\qv}]}$ can be neglected when calculating expectation values. 
Rescaling fermionic fields on the Green's function $g_{\nu\sigma}$ of impurity problem~\eqref{eq:actionimpapp} as $f^{*}_{\kv\nu\sigma}\to{}f^{*}_{\kv\nu\sigma}g^{-1}_{\nu\sigma}$ and $f_{\kv\nu\sigma}\to{}g^{-1}_{\nu\sigma}f_{\kv\nu\sigma}$, and bosonic field on the susceptibility $[\chi_{\omega}]_{\varsigma\varsigma'}$ as $\phi^{*\,\varsigma}_{\qv\omega}\to{}\phi^{*\,\varsigma'}_{\qv\omega}\,[\chi^{\phantom{*}}_{\omega}]^{-1}_{\varsigma'\varsigma}$,
and shifting bosonic variables, the nonlocal part~\eqref{eq:Simp_SM} of the lattice action~\eqref{eq:actionlattapp} transforms to
\begin{align} 
{\cal S}_{\rm DB}
= &-\sum_{\kv,\nu,\sigma} f^{*}_{\kv\nu\sigma}g^{-1}_{\nu\sigma} [\varepsilon^{\phantom{*}}_{\kv}-\Delta^{\phantom{*}}_{\nu\sigma}]^{-1} g^{-1}_{\nu\sigma}f^{\phantom{*}}_{\kv\nu\sigma} 
+\sum_{\kv,\nu,\sigma} \left[ c^{*}_{\kv\nu\sigma} g^{-1}_{\nu\sigma}f^{\phantom{*}}_{\kv\nu\sigma} + 
f^{*}_{\kv\nu\sigma} g^{-1}_{\nu\sigma}c^{\phantom{*}}_{\kv\nu\sigma} \right] +\sum_{\qv,\omega,\varsigma(')} \phi^{*\,\varsigma}_{\qv\omega} \left[\chi^{\phantom{*}}_{\omega}\right]^{-1}_{\varsigma\varsigma'} \rho^{\,\varsigma'}_{\qv\omega} 
\notag\\
&-\frac12\sum_{\qv,\omega,\varsigma(')} \left( \phi^{*\,\varsigma}_{\qv\omega} - j^{*\,\varsigma'}_{\qv\omega} \left[\chi^{\phantom{*}}_{\omega}\right]_{\varsigma'\varsigma} \right)
\left[\chi^{\phantom{*}}_{\omega}\right]^{-1}_{\varsigma\varsigma''} 
\left[V^{\phantom{*}}_{\qv} - \Lambda^{\phantom{*}}_{\omega}\right]^{-1}_{\varsigma''\varsigma'''} \left[\chi^{\phantom{*}}_{\omega}\right]^{-1}_{\varsigma'''\varsigma''''} \left( \phi^{\,\varsigma''''}_{\qv\omega} - \left[\chi^{\phantom{*}}_{\omega}\right]_{\varsigma''''\varsigma'''''}
j^{\,\varsigma'''''}_{\qv\omega}\right) .
\end{align}
Now, the initial degrees of freedom can be integrated out with respect to the impurity action~\eqref{eq:actionimpapp} in the following way
\begin{align}
&\int D[c^{*},c]\,\exp\left\{-\sum_{i}{\cal S}^{\,i}_{\rm imp} - \sum_{\kv,\nu,\sigma} \left[ c^{*}_{\kv\nu\sigma} g^{-1}_{\nu\sigma}f^{\phantom{*}}_{\kv\nu\sigma} + 
f^{*}_{\kv\nu\sigma} g^{-1}_{\nu\sigma}c^{\phantom{*}}_{\kv\nu\sigma} \right] 
-\sum_{\qv,\omega,\varsigma(')} \phi^{*\,\varsigma}_{\qv\omega} \left[\chi^{\phantom{*}}_{\omega}\right]^{-1}_{\varsigma\varsigma'} \rho^{\,\varsigma'}_{\qv\omega} 
\right\} = \notag\\
&{\cal Z}_{\rm imp} \times \exp\left\{ -\sum_{\kv,\nu,\sigma}f^{*}_{\kv\nu\sigma} g^{-1}_{\nu\sigma}f^{\phantom{*}}_{\kv\nu\sigma} 
-\frac12\sum_{\qv,\omega,\varsigma(')} \phi^{*\,\varsigma}_{\qv\omega} \left[\chi^{\phantom{*}}_{\omega}\right]^{-1}_{\varsigma\varsigma'} \phi^{\,\varsigma'}_{\qv\omega} - \tilde{W}[f,\phi]\right\},
\end{align}
where ${\cal Z}_{\rm imp}$ is a partition function of the impurity problem.
Here, the interaction part of the action $\tilde{W}[f,\phi]$ is presented as an infinite series of full vertex functions of impurity problem~\eqref{eq:actionimpapp} as discussed in~\cite{Rubtsov20121320, PhysRevB.93.045107}. The lowest order interaction terms are following 
\begin{align}
\tilde{W}[f,\phi]
&\simeq\sum_{\kv,\kv',\qv}\sum_{\nu,\nu',\omega}\sum_{\sigma('),\varsigma(')}\left(\phi^{*\,\varsigma}_{\qv\omega} 
\gamma^{\,\varsigma}_{\nu\omega} f^{*}_{\kv\nu\sigma} f^{\phantom{*}}_{\kv+\qv,\nu+\omega,\sigma'} 
- \frac14\,\overline{\gamma}^{\,\sigma\sigma'\sigma''\sigma'''}_{\nu\nu'\omega}\, f^{*}_{\kv\nu\sigma}f^{\phantom{*}}_{\kv+\qv,\nu+\omega,\sigma'}f^{*}_{\kv'+\qv,\nu'+\omega,\sigma''} f^{\phantom{*}}_{\kv'\nu'\sigma'''}\right),
\label{eq:lowestint}
\end{align}
where the full three-point vertex function (and its Hermitian conjugate) is defined as
\begin{align}
\label{eq:3vert}
\gamma^{\,\varsigma}_{\nu\omega} &= \sum_{\varsigma'} 
\left[\chi_{\omega}\right]^{-1}_{\varsigma\varsigma'}\av{\rho^{\,\varsigma'}_{\omega}c^{\phantom{*}}_{\nu\sigma}\,c^{*}_{\nu+\omega,\sigma'}}_{\rm imp}g^{-1}_{\nu\sigma}\,g^{-1}_{\nu+\omega,\sigma'} 
= \sum_{\varsigma'} \av{c^{\phantom{*}}_{\nu\sigma}\,c^{*}_{\nu+\omega,\sigma'}\,\rho^{\,\varsigma'}_{\omega}}_{\rm imp} \left[\chi_{\omega}\right]^{-1}_{\varsigma\varsigma'}g^{-1}_{\nu\sigma}
\,g^{-1}_{\nu+\omega,\sigma'}, 
\\
\left[\gamma^{\,\varsigma}_{\nu\omega}\right]^{*} &= \gamma^{\,\varsigma^{*}}_{\nu+\omega,-\omega} = \sum_{\varsigma'} \av{c^{\phantom{*}}_{\nu+\omega,\sigma'}\,c^{*}_{\nu\sigma}\,\rho^{*\,\varsigma'}_{\omega}}_{\rm imp} \left[\chi_{\omega}\right]^{-1}_{\varsigma'\varsigma} g^{-1}_{\nu+\omega,\sigma'}\,g^{-1}_{\nu\sigma}. \notag
\end{align}
The full four-point vertex determined in the particle-hole channel is equal to
\begin{align}
\overline{\gamma}^{\,\sigma\sigma'\sigma''\sigma'''}_{\nu\nu'\omega} =
\av{c^{\phantom{*}}_{\nu\sigma} c^{*}_{\nu+\omega,\sigma'} c^{\phantom{*}}_{\nu'+\omega,\sigma''} c^{*}_{\nu'\sigma'''}}_{\rm c~ imp}\,g^{-1}_{\nu\sigma}\,g^{-1}_{\nu+\omega,\sigma'}\,g^{-1}_{\nu'+\omega,\sigma''}\,g^{-1}_{\nu'\sigma'''}.
\label{eq:4vertapp}
\end{align}
Therefore, the initial lattice problem~\eqref{eq:actionlattapp} transforms to the following {\it dual} action 
\begin{align}
{\cal \tilde{S}}
= &-\sum_{\kv,\nu,\sigma} f^{*}_{\kv\nu\sigma}g^{-1}_{\nu\sigma} [\varepsilon^{\phantom{*}}_{\kv}-\Delta^{\phantom{*}}_{\nu\sigma}]^{-1} g^{-1}_{\nu\sigma}f^{\phantom{*}}_{\kv\nu\sigma} 
+\sum_{\kv,\nu,\sigma}f^{*}_{\kv\nu\sigma} g^{-1}_{\nu\sigma}f^{\phantom{*}}_{\kv\nu\sigma} 
+\frac12\sum_{\qv,\omega,\varsigma(')} \phi^{*\,\varsigma}_{\qv\omega} \left[\chi^{\phantom{*}}_{\omega}\right]^{-1}_{\varsigma\varsigma'} \phi^{\,\varsigma'}_{\qv\omega} + \tilde{W}[f,\phi] \\
&-\frac12\sum_{\qv,\omega,\varsigma(')} \left( \phi^{*\,\varsigma}_{\qv\omega} - j^{*\,\varsigma'}_{\qv\omega} \left[\chi^{\phantom{*}}_{\omega}\right]_{\varsigma'\varsigma} \right)
\left[\chi^{\phantom{*}}_{\omega}\right]^{-1}_{\varsigma\varsigma''} 
\left[V^{\phantom{*}}_{\qv} - \Lambda^{\phantom{*}}_{\omega}\right]^{-1}_{\varsigma''\varsigma'''} \left[\chi^{\phantom{*}}_{\omega}\right]^{-1}_{\varsigma'''\varsigma''''} \left( \phi^{\,\varsigma''''}_{\qv\omega} - \left[\chi^{\phantom{*}}_{\omega}\right]_{\varsigma''''\varsigma'''''}
j^{\,\varsigma'''''}_{\qv\omega}\right). \notag
\end{align}

In order to come back to the original bosonic variables, one can perform the third Hubbard-Stratonovich transformation as 
\begin{align}
&\exp\left\{\frac12\sum_{\qv,\omega,\varsigma(')} \left( \phi^{*\,\varsigma}_{\qv\omega} - j^{*\,\varsigma'}_{\qv\omega} \left[\chi^{\phantom{*}}_{\omega}\right]_{\varsigma'\varsigma} \right)
\left[\chi^{\phantom{*}}_{\omega}\right]^{-1}_{\varsigma\varsigma''} 
\left[V^{\phantom{*}}_{\qv} - \Lambda^{\phantom{*}}_{\omega}\right]^{-1}_{\varsigma''\varsigma'''} \left[\chi^{\phantom{*}}_{\omega}\right]^{-1}_{\varsigma'''\varsigma''''} \left( \phi^{\,\varsigma''''}_{\qv\omega} - \left[\chi^{\phantom{*}}_{\omega}\right]_{\varsigma''''\varsigma'''''}
j^{\,\varsigma'''''}_{\qv\omega}\right)\right\} = \\
&D_{\bar\rho}
\int D[\bar\rho]\exp\left\{-\sum_{\qv,\omega,\varsigma(')}\left(\frac12\, \bar\rho^{*\,\varsigma}_{\qv\omega}\,
\left[V^{\phantom{*}}_{\qv} - \Lambda^{\phantom{*}}_{\omega} \right]_{\varsigma\varsigma'}\bar\rho^{\,\varsigma'}_{\qv\omega} 
- \phi^{*\,\varsigma}_{\qv\omega} \left[\chi^{\phantom{*}}_{\omega}\right]^{-1}_{\varsigma\varsigma'} \bar\rho^{\,\varsigma'}_{\qv\omega} 
+ j^{\,*\,\varsigma}_{\qv\omega} \bar\rho^{\,\varsigma}_{\qv\omega} \right)\right\}. \notag
\end{align}
Comparing this expression to the Eq.~\ref{eq:Simp_SM}, one can see that sources $j^{*\,\varsigma}_{\qv\omega}$ introduced for the initial degrees of freedom $\rho^{\,\varsigma}_{\qv\omega}$ are also the sources for new bosonic fields $\bar\rho^{\,\varsigma}_{\qv\omega}$. Therefore, fields $\bar\rho^{\,\varsigma}_{\qv\omega}$ indeed represent initial degrees of freedom and have the same physical meaning as original {\it composite} bosonic variables $\rho^{\,\varsigma}_{\qv\omega}=\sum_{\kv\nu\sigma\sigma'}c^{*}_{\kv\nu\sigma} \sigma^{\varsigma}_{\sigma\sigma'} c^{\phantom{*}}_{\kv+\qv,\nu+\omega,\sigma'}-\av{n^{\,\varsigma}_{\qv\omega}}$ of the lattice problem~\eqref{eq:actionlattapp}. Nevertheless, $\bar\rho^{\,\varsigma}_{\qv\omega}$ can now be treated as {\it elementary} bosonic fields that have a well-defined propagator, since they are introduced as a decoupling fields of dual degrees of freedom $\phi^{\,\varsigma}_{\qv\omega}$ and therefore, independent on fermionic variables $c^{*}_{\kv\nu\sigma}\,(c_{\kv\nu\sigma})$. Taking sources to zero and replacing $\bar\rho^{\,\varsigma}_{\qv\omega}$ by $\rho^{\,\varsigma}_{\qv\omega}$, dual bosonic fields can be integrated out with respect to the Gaussian bosonic part of the dual action as
\begin{align}
&\int D[\phi^{*},\phi]\,\exp\left\{-\sum_{\qv,\omega,\varsigma(')} \left(\frac12\,\phi^{*\,\varsigma}_{\qv\omega} \left[\chi^{\phantom{*}}_{\omega}\right]^{-1}_{\varsigma\varsigma'} \phi^{\,\varsigma'}_{\qv\omega} 
- \phi^{*\,\varsigma}_{\qv\omega} \left[\chi^{\phantom{*}}_{\omega}\right]^{-1}_{\varsigma\varsigma'} \bar\rho^{\,\varsigma'}_{\qv\omega} \right) - \tilde{W}[f,\phi]\right\}
= 
{\cal Z}_{\phi} \times \exp\left\{\frac12\sum_{\qv,\omega,\varsigma(')} \rho^{*\,\varsigma}_{\qv\omega} \left[\chi^{\phantom{*}}_{\omega}\right]^{-1}_{\varsigma\varsigma'} \rho^{\,\varsigma'}_{\qv\omega} - W[f,\rho] \right\},
\label{eq:integrationphi}
\end{align}
where ${\cal Z}_{\phi}$ is a partition function of the Gaussian part of the bosonic action.
Here we restrict ourselves to the lowest order interaction terms of $\tilde{W}[f,\phi]$ shown in Eq.~\ref{eq:lowestint}. Then, the integration of dual bosonic fields in Eq.~\ref{eq:integrationphi} simplifies and $W[f,\rho]$ keeps an efficient dual form of $\tilde{W}[f,\phi]$~\eqref{eq:lowestint} with replacement of bosonic variables $\phi^{\,\varsigma}\to\bar\rho^{\,\varsigma}$. Also the four-point vertex becomes irreducible with respect to the full local bosonic propagator $\chi_{\omega}$, as can be seen from the works of~\cite{doi:10.1142/S0217979200002430, DUPUIS2001617, 2001cond.mat.DUPUIS}, while the three-point vertex $\gamma_{\nu\omega}$ remains invariant
\begin{align}
W[f,\rho]
=\sum_{\kv,\kv',\qv}\sum_{\nu,\nu',\omega}\sum_{\varsigma(')}\left(\rho^{*\,\varsigma}_{\qv\omega} 
\gamma^{\,\varsigma}_{\nu\omega} f^{*}_{\kv\nu\sigma} f^{\phantom{*}}_{\kv+\qv,\nu+\omega,\sigma'} 
- \left[\overline{\gamma} - \theta\right]^{\,\varsigma\varsigma'}_{\nu\nu'\omega} f^{*}_{\kv\nu\sigma}f^{\phantom{*}}_{\kv+\qv,\nu+\omega,\sigma'}f^{*}_{\kv'+\qv,\nu'+\omega,\sigma''} f^{\phantom{*}}_{\kv'\nu'\sigma'''}\right).
\label{eq:Wfull}
\end{align}
Here,
\begin{align}
\theta^{\,\varsigma\varsigma'}_{\nu\nu'\omega} = - \gamma^{\,\varsigma}_{\nu\omega}\,\left[\chi_{\omega}\right]_{\varsigma\varsigma'} \left[\gamma^{\,\varsigma'}_{\nu'\omega}\right]^{*}
\label{eq:Thetaapp}
\end{align}
is the full reducible bosonic contribution to the full local four-point vertex $\overline{\gamma}^{\,\varsigma\varsigma'}_{\nu\nu'\omega}$ introduced in~\cite{PhysRevB.94.205110} and spin labels $\sigma, \sigma', \sigma'', \sigma'''$ are fixed by the channel indices $\varsigma, \varsigma'$.
Therefore, the problem transforms to the following action of an effective $s$-$d$ model 
\begin{align}
{\cal S}_{s\text{-}d} = &-\sum_{\kv,\nu,\sigma} f^{*}_{{\bf k}\nu\sigma}\tilde{G}^{-1}_{0}f^{\phantom{*}}_{{\bf k}\nu\sigma} 
-\frac12\sum_{\qv,\omega,\varsigma(')} \rho^{*\,\varsigma}_{\qv\omega}\left[X^{\phantom{*}}_{\rm E}\right]^{-1}_{\varsigma\varsigma'}\rho^{\,\varsigma'}_{\qv\omega} + W[f,\rho],
\label{eq:actionSDapp}
\end{align}
where $[X_{\rm E}]_{\varsigma\varsigma'} = \left[\chi^{-1}_{\omega} + \Lambda^{\phantom{1}}_{\omega} - V^{\phantom{1}}_{\qv} \right]^{-1}_{\varsigma\varsigma'}$ is the susceptibility of the extended dynamical mean-field theory.
As it is shown below, when the three-point vertex function $\gamma'_{\nu\omega}$ of impurity problem that connects two fermionic propagators and interaction is close to unity~\eqref{eq:vertex_unity}, the main contribution to the local four-point vertex is given by the full reducible bosonic contribution, i.e. $\overline{\gamma}\simeq\theta$, or diagrammatically
\begin{align}
\includegraphics[width=0.15\linewidth]{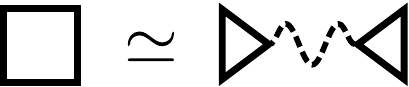}~.
\label{eq:PiDualApp}
\end{align}
Here, the dotted wave line depicts full local bosonic propagator and the minus sign in Eq.~\ref{eq:Thetaapp} appears due to Feinman rules~\cite{PhysRevB.90.235135}. Then, the interaction part of the action~\eqref{eq:actionSDapp} takes the most simple form that contains only three-point vertex functions
\begin{align}
W'[f,\rho]
\simeq\sum_{\kv,\qv}\sum_{\nu,\omega}\sum_{\sigma('),\varsigma(')}
\rho^{*\,\varsigma}_{\qv\omega} 
\gamma^{\,\varsigma}_{\nu\omega} f^{*}_{\kv\nu\sigma} f^{\phantom{*}}_{\kv+\qv,\nu+\omega,\sigma'} 
.
\label{eq:Wsimple}
\end{align}

\subsection*{Transformation of spin basis}
Let us consider an effective impurity model in the spin-polarized case. For easier description, one can transform spin variables from the ${\bf s}=\{x,y,z\}$ to the ${\bf s}=\{+,-,z\}$ basis as $S^{\pm}=(\rho^{x}\pm{}i\rho^{y})/2$. In the spin-polarized case fluctuations in the charge and spin $z$ channels are yet entangled, but the $\pm$ spin channel can be separated in the collinear case. Thus, for a correct account for spin fluctuations, one may consider correlations only in the $\pm$ spin channel and the contribution of the $z$ channel to the exchange interaction can be later restored from the symmetry arguments. In is worth mentioning that the transformation $\{x,y\}\to\{+,-\}$ is very useful for calculation of physical observables, since it diagonalizes the spin susceptibility. Nevertheless, one has to remember that operators $S^{+}$ and $S^{-}$ are not Hermitian. Therefore, components of bosonic operator in matrix representation in the old and new basis are defined as
\begin{align}
\hat\rho_{\qv\omega} = 
\begin{pmatrix}
\rho^{\,x}_{\qv\omega} \\
\rho^{\,y}_{\qv\omega} \\ 
\end{pmatrix}; ~~
\hat{S}_{\qv\omega} = 
\begin{pmatrix}
S^{+}_{\qv\omega} \\
S^{-}_{\qv\omega} \\ 
\end{pmatrix}; ~~
\hat\rho^{*}_{\qv\omega} = 
\Big(\rho^{\,x}_{-\qv,-\omega},~ \rho^{\,y}_{-\qv,-\omega}\Big); ~~
\hat{S}^{*}_{\qv\omega} = 
\Big(S^{-}_{-\qv,-\omega},~ S^{+}_{-\qv,-\omega}\Big).
\end{align}
Connection between these bases can be obtained using the following matrix transformation
\begin{align}
\hat{S}^{*}_{\qv\omega} = \hat{\rho}^{*}_{\qv\omega} \times \hat{A},
~~~~\text{or}~~~~
\Big(S^{-}_{-\qv,-\omega},~ S^{+}_{-\qv,-\omega}\Big) = 
\Big(\rho^{\,x}_{-\qv,-\omega},~ \rho^{\,y}_{-\qv,-\omega}\Big) \times
\begin{pmatrix}
\frac{1}{2} & \frac{1}{2} \\
\frac{-i}{2} & \frac{i}{2} \\ 
\end{pmatrix}
\end{align}
and
\begin{align}
\hat{S}_{\qv\omega} = \hat{B} \times \hat\rho_{\qv\omega},
~~~~\text{or}~~~~
\begin{pmatrix}
S^{+}_{\qv\omega} \\
S^{-}_{\qv\omega} \\ 
\end{pmatrix} = 
\begin{pmatrix}
\frac{1}{2} & \frac{i}{2} \\
\frac{1}{2} & \frac{-i}{2} \\ 
\end{pmatrix}
\times
\begin{pmatrix}
\rho^{\,x}_{\qv\omega} \\
\rho^{\,y}_{\qv\omega} \\ 
\end{pmatrix}.
\end{align}
Then, all matrices $\hat{M}_{xy}$ involved in above derivations can also be transformed to the new basis $\hat{M}_{\pm}$ as
\begin{align}
\hat{M}_{\pm} = \hat{A}^{-1} \times \hat{M}_{xy} \times \hat{B}^{-1}.
\end{align}
In particular, the matrix form of the nonlocal interaction $[V_{\qv}]_{\rm \varsigma\varsigma'}$ remains diagonal
\begin{align}
\left[V_{\qv}\right]_{\pm} = 
\hat{A}^{-1} \times
\begin{pmatrix}
-J^{\rm d}_{\qv}/2 & 0 \\
0 & -J^{\rm d}_{\qv}/2 \\ 
\end{pmatrix}
\times \hat{B}^{-1} = 
\begin{pmatrix}
-J^{\rm d}_{\qv} & 0 \\
0 & -J^{\rm d}_{\qv} \\ 
\end{pmatrix}
\end{align}
and inverse susceptibility is transformed to a diagonal form as
\begin{align}
\left[\chi_{\omega}\right]_{\pm}^{-1} = 
\hat{A}^{-1} \times
\begin{pmatrix}
\chi^{xx}_{\omega} & \chi^{xy}_{\omega} \\
\chi^{yx}_{\omega} & \chi^{yy}_{\omega} \\ 
\end{pmatrix}
^{-1}
\times \hat{B}^{-1} = 
\frac{1}{\chi^{xx}_{\omega}\chi^{yy}_{\omega} - \chi^{xy}_{\omega}\chi^{yx}_{\omega}}
\begin{pmatrix}
\chi^{xx}_{\omega}+\chi^{yy}_{\omega} + i\chi^{xy}_{\omega} - i\chi^{yx}_{\omega}  & \chi^{xx}_{\omega}-\chi^{yy}_{\omega} + i\chi^{xy}_{\omega} + i\chi^{yx}_{\omega} \\
-\chi^{xx}_{\omega}+\chi^{yy}_{\omega} + i\chi^{xy}_{\omega} + i\chi^{yx}_{\omega} & \chi^{xx}_{\omega}+\chi^{yy}_{\omega} - i\chi^{xy}_{\omega} + i\chi^{yx}_{\omega} \\ 
\end{pmatrix}.
\end{align}
Defining $\chi^{+-}_{\omega} = -\av{S^{+}_{\omega}\,S^{-}_{-\omega}} = \frac14(\chi^{xx}_{\omega}+\chi^{yy}_{\omega} - i\chi^{xy}_{\omega} + i\chi^{yx}_{\omega})$ and $\chi^{-+}_{\omega} = -\av{S^{-}_{\omega}\,S^{+}_{-\omega}} = \frac14(\chi^{xx}_{\omega}+\chi^{yy}_{\omega} + i\chi^{xy}_{\omega} - i\chi^{yx}_{\omega})$, and taking into account that $\chi^{xx}_{\omega} = \chi^{yy}_{\omega}$ and $\chi^{xy}_{\omega} = -\chi^{yx}_{\omega}$, one gets that
$
\chi^{xx}_{\omega}\chi^{yy}_{\omega} - \chi^{xy}_{\omega}\chi^{yx}_{\omega} = 4\,\chi^{+-}_{\omega} \chi^{-+}_{\omega}
$
and 
\begin{align}
\left[\chi_{\omega}\right]_{\pm}^{-1}  = 
\begin{pmatrix}
[\chi^{+-}_{\omega}]^{-1}  & 0 \\
0 & [\chi^{-+}_{\omega}]^{-1} \\ 
\end{pmatrix}.
\end{align}

\subsection*{Magnetic susceptibility}

In order to obtain the effective problem written in terms of bosonic degrees of freedom only, one can integrate out dual fermionic degrees of freedom from the Eq.~\ref{eq:actionSDapp}.
Taking into account transformation of the spin basis presented above, the spin $\pm$ part of the effective action reads
\begin{align}
{\cal S}_{\rm spin}
&= -\frac12\sum_{\qv,\omega} S^{\,-}_{\qv\omega} 
\left[X_{\qv\omega}^{\,-+}\right]^{-1} \, S^{\,+}_{-\qv,-\omega}
-\frac12\sum_{\qv,\omega} S^{\,+}_{\qv\omega} 
\left[X_{\qv\omega}^{\,+-}\right]^{-1} \, S^{\,-}_{-\qv,-\omega}
\label{eq:Spmapp}
\end{align}
The first approximation for the spin susceptibility $X^{-+}_{\qv\omega}$ can be obtained after expanding the simplified form of interaction $W[f,\rho]$ given by Eq.~\ref{eq:Wsimple} up to the second order with respect to bosonic fields $\rho$ in the expression for the partition function of the action~\eqref{eq:actionSDapp}. This results in 
\begin{align}
\left[X^{(2)}_{\qv\omega}\right]^{-1}
&= J^{\rm d}_{\bf q} + \Lambda^{\phantom{1}}_{\omega} + \chi^{-1}_{\omega} - \tilde{\Pi}^{(2)}_{\qv\omega},
\label{eq:X1pmapp}
\end{align}
where
\begin{align}
\tilde{\Pi}^{(2)}_{\qv\omega}=\sum_{\kv,\nu}
\gamma^{\,-}_{\nu+\omega,-\omega}\,\tilde{G}^{\phantom{2}}_{\kv+\qv,\nu+\omega\uparrow}\tilde{G}^{\phantom{2}}_{\kv\nu\downarrow}\,\gamma^{\,+}_{\nu,\omega}
\end{align}
is the second order polarization function and $\chi^{\phantom{+}}_{\omega}=\chi^{-+}_{\omega}$ and $\Lambda^{\phantom{+}}_{\omega}=\Lambda^{-+}_{\omega}$ are the spin susceptibility and bosonic hybridization function of impurity problem, respectively. Hereinafter, $\pm$ spin labels are omitted for simplicity wherever they are not crucial for understanding. The three-point vertex functions in the spin channel are defined as in Eq.~\ref{eq:3vert}, or explicitly as
\begin{align}
\label{eq:3vertpmapp}
\gamma^{\,+}_{\nu,\omega} &=   \av{c^{\phantom{*}}_{\nu\downarrow}\,c^{*}_{\nu+\omega\uparrow}\,S^{-}_{\omega}}_{\rm imp} \left[\chi^{-+}_{\omega}\right]^{-1}g^{-1}_{\nu\downarrow}
\,g^{-1}_{\nu+\omega\uparrow}, \\
\gamma^{\,-}_{\nu+\omega,-\omega} &=  
\av{S^{+}_{-\omega}\,c^{\phantom{*}}_{\nu+\omega\uparrow}\,c^{*}_{\nu\downarrow}}_{\rm imp} \left[\chi^{-+}_{\omega}\right]^{-1} g^{-1}_{\nu+\omega\uparrow}\,g^{-1}_{\nu\downarrow}. \notag
\end{align}
The more accurate approximation for the spin susceptibility can be found when expanding the full form of interaction  $W[f,\rho]$ given by Eq.~\ref{eq:Wfull} up to the second order with respect to bosonic fields $\rho$ as previously. Using the ladder approximation, one gets 
\begin{align}
\left[X^{\rm ladd}_{\qv\omega}\right]^{-1} =
J^{\rm d}_{\bf q} + \Lambda^{\phantom{1}}_{\omega} + \chi^{-1}_{\omega} - \Pi^{\rm ladd}_{\qv\omega},
\label{eq:Pidirectapp}
\end{align}
where the polarization function $\Pi^{\rm ladd}_{\qv\omega}$ expressed in the matrix form in the space of fermionic frequencies $\nu,\nu'$ reads
\begin{align}
\Pi^{\rm ladd}_{\qv\omega} = \Tr\left\{ \hat\gamma^{-}_{\omega}\,\hat{\tilde{X}}^{0}_{\qv\omega}\left[ I + \left(\hat{\overline{\gamma}}^{\phantom{+}}_{\omega}-\hat\theta^{\phantom{+}}_{\omega}\right) \hat{\tilde{X}}^{0}_{\qv\omega} \right]^{-1} \hat\gamma^{+}_{\omega} \right\}.
\label{eq:Pilatladdapp}
\end{align} 
Here, $I$ is the identity matrix in the same space. Multiplication and inversion should be understood as a standard matrix operations. For simplicity, we omit the fermionic indices  wherever they are not crucial for understanding. 
The trace is taken over the external fermionic indices.  
Matrix elements of the bare dual spin susceptibility $\tilde{X}^{0}_{\qv\omega}$ and three-point vertex function $\gamma_{\omega}$ are defined as $\tilde{X}^{0}_{\qv\omega;\,\nu\nu'} = \sum_{\kv}\tilde{G}_{\kv+\qv,\nu+\omega\uparrow}\tilde{G}_{\kv\nu\downarrow}\,\delta_{\nu\nu'}$ and $\gamma^{\pm}_{\omega; \nu\nu'} = \gamma^{\pm}_{\nu\omega}\,\delta_{\nu\nu'}$, where $\gamma^{\pm}_{\nu\omega}$ are defined in Eq.~\ref{eq:3vertpmapp}. The four-point vertex functions $\overline{\gamma}_{\nu\nu'\omega}$ and $\theta_{\nu\nu'\omega}$ in the $\pm$ spin channel are defined above in Eqs.~\ref{eq:4vertapp} and~\eqref{eq:Thetaapp}, or explicitly as
\begin{align}
\overline{\gamma}^{\phantom{1}}_{\nu\nu'\omega} &= \overline{\gamma}^{\,\downarrow\uparrow\uparrow\downarrow}_{\nu\nu'\omega} =
\av{c^{\phantom{*}}_{\nu\downarrow} c^{*}_{\nu+\omega\uparrow} c^{\phantom{*}}_{\nu'+\omega\uparrow} c^{*}_{\nu'\downarrow}}_{\rm c~ imp}\,g^{-1}_{\nu\downarrow}\,g^{-1}_{\nu+\omega\uparrow}\,g^{-1}_{\nu'+\omega\uparrow}\,g^{-1}_{\nu'\downarrow}, \label{eq:4vertexplapp}\\
\theta^{\phantom{1}}_{\nu\nu'\omega} &= - \gamma^{\,+}_{\nu\omega}\,\chi_{\omega} \gamma^{\,-}_{\nu'+\omega,-\omega}.\label{eq:33vertexplapp}
\end{align}
Substituting the above expressions to the Eq.~\ref{eq:Pidirectapp}, one recovers conserving result for the spin susceptibility provided by the ladder DB approach~\cite{PhysRevB.93.045107} in the case of the constant bosonic hybridization function $\Lambda$~\cite{KrienF}
\begin{align}
\left[X^{\rm ladd}_{\qv\omega}\right]^{-1} = J^{\rm d}_{\qv} + \Lambda + \left[X^{\rm DMFT}_{\qv\omega}\right]^{-1}.
\label{eq:X_DBapp}
\end{align}
Here, 
\begin{align}
X^{\rm DMFT}_{\qv\omega}=\chi^{\phantom{2}}_{\omega}+\chi^{\phantom{2}}_{\omega}\tilde{\Pi}^{\rm ladd}_{\qv\omega}\chi^{\phantom{2}}_{\omega}
\label{eq:XDMFTapp}
\end{align}
and $\tilde{\Pi}^{\rm ladd}_{\qv\omega}$ is the dual polarization function in the ladder form~\cite{PhysRevB.90.235105} given by the following matrix form in the space of fermionic frequencies $\nu,\nu'$
\begin{align}
\tilde{\Pi}^{\rm ladd}_{\qv\omega} = \Tr\left\{ \hat\gamma^{-}_{\omega}\,\hat{\tilde{X}}^{0}_{\qv\omega}\left[ I + \hat{\overline{\gamma}}^{\phantom{+}}_{\omega} \hat{\tilde{X}}^{0}_{\qv\omega} \right]^{-1} \hat\gamma^{+}_{\omega} \right\}.
\label{eq:Piladdapp}
\end{align} 
As it was already noted in~\cite{PhysRevB.94.205110}, the difference between the lattice~\eqref{eq:Pilatladdapp} and dual~\eqref{eq:Piladdapp} polarization functions is that the first one is irreducible with respect to the (local and nonlocal parts of) EDMFT susceptibility $X_{\rm E}$, while the dual one is irreducible only with respect to the bare dual susceptibility, which is identically equal to the nonlocal part of $X_{\rm E}$.

Expression for the spin susceptibility~\eqref{eq:X_DBapp} can be rewritten in the more convenient way. 
For this reason one can define the two-particle irreducible (2PI) vertex function in the $\pm$ spin channel as 
\begin{align}
\hat{\overline{\gamma}}^{\rm 2PI}_{\omega} = \hat{\overline{\gamma}}^{\phantom{1}}_{\omega} \left[I - \hat\chi^{0}_{\omega}\hat{\overline{\gamma}}^{\phantom{1}}_{\omega}\right]^{-1},
\end{align}
where the matrix elements of the bare local spin susceptibility are $\chi^{0}_{\omega;\,\nu\nu'} = g_{\nu+\omega\uparrow}g_{\nu\downarrow}\,\delta_{\nu\nu'}$. 
Then, the spin susceptibility of the impurity problem can be expressed as
\begin{align}
\chi_{\omega} = -\av{S^{-}_{\omega}\,S^{+}_{-\omega}} = \Tr\left\{\hat\chi^{0}_{\omega} - \hat\chi^{0}_{\omega}\,\hat{\overline{\gamma}}_{\omega} \,\hat\chi^{0}_{\omega} \right\} = \Tr\left\{\hat\chi^{0}_{\omega} \left[I+\hat{\overline{\gamma}}^{\rm 2PI}_{\omega}\hat\chi^{0}_{\omega}\right]^{-1}\right\}.
\end{align} 
Rewriting the relation for the dual polarization function $\tilde{\Pi}^{\rm ladd}_{\qv\omega}$~\eqref{eq:Piladdapp} through the 2PI vertex function and using the exact relation between the three- and four-point vertex functions of impurity problem
\begin{align}
\label{eq:3-4app}
\gamma^{+}_{\nu\omega} &=  
\av{c^{\phantom{*}}_{\nu\downarrow}\,c^{*}_{\nu+\omega\uparrow}\,S^{-}_{\omega}}_{\rm imp} \chi^{-1}_{\omega}g^{-1}_{\nu\downarrow}
\,g^{-1}_{\nu+\omega\uparrow} 
= \sum_{\nu'} \av{c^{\phantom{*}}_{\nu\downarrow}\,c^{*}_{\nu+\omega\uparrow}\,
c^{*}_{\nu'\downarrow}\,c^{\phantom{*}}_{\nu'+\omega,\uparrow}}_{\rm imp} \chi^{-1}_{\omega}g^{-1}_{\nu\downarrow}
\,g^{-1}_{\nu+\omega\uparrow} \\
&= \sum_{\nu'}\left\{\delta_{\nu,\nu'} -  \overline\gamma^{\phantom{1}}_{\nu,\nu',\omega}\,g_{\nu'+\omega\uparrow}\,g_{\nu'\downarrow} \right\} \chi^{-1}_{\omega}
= \sum_{\nu'} \left[\delta_{\nu\nu'} + \overline\gamma^{\rm 2PI}_{\nu,\nu',\omega}\, g_{\nu'+\omega\uparrow}g_{\nu'\downarrow}\right]^{-1}
\chi^{-1}_{\omega}, \notag 
\end{align}
and the fact that in the case of zero dual self energy $\tilde{\Sigma}_{\kv\nu}=0$ the following relation holds 
\begin{align}
\tilde{X}^{0}_{\qv\omega;\,\nu\nu'} + \chi^{0}_{\omega;\,\nu\nu'} = X^{0}_{\qv\omega;\,\nu\nu'} = \sum_{\kv}G_{\kv+\qv,\nu+\omega\uparrow}G_{\kv\nu\downarrow}\,\delta_{\nu\nu'},
\end{align}
one finds that
\begin{align}
X^{\rm DMFT}_{\qv\omega} = \chi^{\phantom{1}}_{\omega} + \chi^{\phantom{1}}_{\omega}\tilde\Pi^{\rm ladd}_{\qv\omega} \chi^{\phantom{1}}_{\omega} = 
\Tr\left\{\hat{X}^{0}_{\qv\omega} \left[I + \hat{\overline{\gamma}}^{\,\rm 2PI}_{\omega}  \hat{X}^{0}_{\qv\omega} \right]^{-1}\right\}
\end{align}
is the DMFT-like~\cite{PhysRevLett.62.324, RevModPhys.68.13}
susceptibility written  in terms of the 2PI vertex functions of impurity model and lattice Green's functions. 
Therefore, the spin susceptibility~\eqref{eq:X_DBapp} derived within the ladder Dual Boson approach~\cite{Rubtsov20121320} can be rewritten as
\begin{align}
X^{\rm ladd}_{\qv\omega} = \Tr\left\{\hat{X}^{0}_{\qv\omega} \left[ I + \left(\hat{\overline{\gamma}}^{\rm 2PI}_{\omega} + I\left[J^{\rm D}_{\bf q} + \Lambda\right]\right) \hat{X}^{0}_{\qv\omega} \right]^{-1}\right\}.
\label{eq:Xladdapp}
\end{align}

\subsection*{Classical Heisenberg Hamiltonian} 
In order to map the initial problem onto a classical Heisenberg Hamiltonian the
spin variables $S^{\pm}_{\qv\omega}$ in Eq.~(\ref{eq:Spmapp}) have to be replaced by the classical vectors ${\bf S}_{\qv}$ of the length $\av{S_{z}}$.
Then, an effective exchange interaction $J_{\qv}$ can be defined as a nonlocal part of the inverse susceptibility at zero bosonic frequency~\cite{KL2000}.
After all, the action~\eqref{eq:Spmapp} maps on an effective Heisenberg Hamiltonian 
\begin{align}
H_{\rm spin} = - \sum_{\qv} J_{\qv}\,{\bf S}_{\qv}\,{\bf S}_{-\qv},
\end{align}
where the contribution from the $z$ spin channel is restored from the requirement of rotational invariance. Here, the effective exchange interaction obtained from the simplified form of magnetic susceptibility~\eqref{eq:X1pmapp} is
\begin{align}
J_{\qv} = J^{\rm d}_{\qv} - \sum_{\kv,\nu}
\gamma^{\,-}_{\nu,\omega=0}\,\tilde{G}^{\phantom{2}}_{\kv+\qv,\nu\uparrow}\tilde{G}^{\phantom{2}}_{\kv\nu\downarrow}\,\gamma^{\,+}_{\nu,\omega=0}.
\label{eq:Jfinalapp}
\end{align}
and the exchange interaction in the ladder approximation obtained from the Eq.~\ref{eq:Xladdapp} reads
\begin{align}
J^{\phantom{d}}_{\qv} = J^{\rm d}_{\qv} - \tilde{\Pi}^{\rm ladd}_{\qv,\omega=0}\left[1+\chi^{\phantom{d}}_{\omega=0}\,\tilde{\Pi}^{\rm ladd}_{\qv,\omega=0}\right]^{-1}.
\label{eq:Jfinalapp2}
\end{align}

\subsection*{Ward identity for the vertex function of impurity model}

When the system exhibits mostly bosonic fluctuation, one can expect that local vertex functions of impurity problem are mostly described by the bosonic frequency $\omega$, while the dependence on fermionic frequencies $\nu,\nu'$ can be averaged. In order to account for single electronic degrees of freedom correctly, the averaging procedure over the fermionic frequencies is carried out using Ward identity for the two-particle irreducible four-point vertex function of the impurity problem~\cite{KrienF} as
\begin{align}
\label{eq:zamena}
\Sigma_{\nu+\omega\uparrow}-\Sigma_{\nu\downarrow} = -\sum_{\nu''} \overline\gamma^{\rm 2PI}_{\nu,\nu'',\omega} \left(g_{\nu''+\omega\uparrow}-g_{\nu''\downarrow}\right) \simeq -\,\overline\gamma^{\rm 2PI}_{\nu,\av{\nu''},\omega}
\left\{\sum_{\nu''} \left(g_{\nu''+\omega\uparrow}-g_{\nu''\downarrow}\right)\right\} 
= - 2\,\overline\gamma^{\rm 2PI}_{\nu,\av{\nu''},\omega}\,\av{S^{z}}.
\end{align}
Then, one can approximate the two-particle irreducible vertex function as
\begin{align}
\overline\gamma^{\rm 2PI}_{\nu,\nu'',\omega} \simeq \overline\gamma^{\rm 2PI}_{\nu,\av{\nu''},\omega} = - \frac{\Sigma_{\nu+\omega\uparrow}-\Sigma_{\nu\downarrow}}{2\av{S^{z}}} = 
-\delta\Sigma_{\nu\omega}.
\label{eq:2piapproxapp}
\end{align}
The three-point vertex function~\eqref{eq:3-4app} is then simplified as
\begin{align}
\label{eq:3approxWapp}
\gamma^{+}_{\nu\omega} 
&= \sum_{\nu'\nu''}\left\{\delta_{\nu,\nu'} -  \overline\gamma^{\rm 2PI}_{\nu,\nu'',\omega}\left[\delta_{\nu''\nu'} + g_{\nu''+\omega\uparrow}g_{\nu''\downarrow}\overline\gamma^{\rm 2PI}_{\nu'',\nu',\omega}\right]^{-1}
g_{\nu'+\omega\uparrow}\,g_{\nu'\downarrow} \right\} \chi^{-1}_{\omega} \\
&\simeq  \left\{1 - \overline\gamma^{\rm 2PI}_{\nu,\av{\nu''},\omega}\sum_{\nu'\nu''}\left\{\left[\delta_{\nu''\nu'} + g_{\nu''+\omega\uparrow}g_{\nu''\downarrow}\overline\gamma^{\rm 2PI}_{\nu'',\nu',\omega}\right]^{-1}
g_{\nu'+\omega\uparrow}\,g_{\nu'\downarrow}\right\} \right\}\chi^{-1}_{\omega} \notag\\
&=  \left\{1 - \overline\gamma^{\rm 2PI}_{\nu,\av{\nu''},\omega}\,\chi^{\phantom{1}}_{\omega} \right\} \chi^{-1}_{\omega}
= \chi^{-1}_{\omega} +\delta\Sigma_{\nu\omega} \simeq \chi^{\rm 0~-1}_{\omega}, \notag
\end{align}
where $\chi^{\rm 0}_{\omega} = \sum_{\nu}g_{\nu+\omega\uparrow}\,g_{\nu\downarrow}$.
Similarly, one can show that 
\begin{align}
\gamma^{-}_{\nu+\omega,-\omega} 
\simeq \chi^{-1}_{\omega} + \delta\Sigma_{\nu\omega} \simeq \chi^{\rm 0~-1}_{\omega}
\label{eq:3vertexapp}
\end{align}
and the magnetic susceptibility~\eqref{eq:X1pmapp} can be written as
\begin{align}
\Big[X^{(2)}_{\qv\omega}\Big]^{-1} &= J^{\rm d}_{\bf q} + \Lambda^{\phantom{1}}_{\omega} + \chi^{-1}_{\omega} - \sum_{\kv,\nu}\left(\chi^{-1}_{\omega} + \delta\Sigma_{\nu\omega}\right)\,\tilde{G}_{\kv+\qv,\nu+\omega,\uparrow}\,\tilde{G}_{\kv,\nu,\downarrow}\,\left(\chi^{-1}_{\omega} + \delta\Sigma_{\nu\omega}\right)\\
&= J^{\rm d}_{\bf q} + \Lambda^{\phantom{1}}_{\omega} + \chi^{-1}_{\omega} - \chi^{\rm 0~-1}_{\omega}\,\tilde{X}^{0}_{\qv\omega}\,\chi^{\rm 0~-1}_{\omega}.
\label{eq:X2approxapp}
\end{align}
Here, we also introduce $\tilde{X}^{0}_{\qv\omega} = \sum_{\kv,\nu}\tilde{G}_{\kv+\qv,\nu+\omega\uparrow}\,\tilde{G}_{\kv,\nu\downarrow}$.

The ladder form of the magnetic susceptibility~\eqref{eq:Xladdapp} can also be simplified. Taking into account that the last approximation in Eq.~\ref{eq:3vertexapp} is nothing else than averaging of the 2PI four-point vertex function~\eqref{eq:2piapproxapp} over the second fermionic frequency 
\begin{align}
\overline\gamma^{\rm 2PI}_{\nu\av{\nu'}\omega}\simeq\overline\gamma^{\rm 2PI}_{\av{\nu\nu'}\omega}=\chi^{-1}_{\omega}-\chi^{0~-1}_{\omega},
\label{eq:gamma2PIapp}
\end{align}
one gets the following RPA-like approximation for the magnetic susceptibility in the ladder approximation~\eqref{eq:Xladdapp}
\begin{align}
X^{\rm ladd}_{\qv\omega} = X^{0}_{\qv\omega} \left[1+\left(\overline\gamma^{\rm 2PI}_{\av{\nu\nu'}\omega} + \Lambda + J^{\rm d}_{\bf q}\right)X^{0}_{\qv\omega}\right]^{-1},
\label{eq:X_DMFTapp}
\end{align}
where the bare lattice magnetic susceptibility $X^{0}_{\qv\omega} = \sum_{\kv\nu}G_{\kv+\qv,\nu+\omega\uparrow}G_{\kv\nu\downarrow}$ was introduced.

As it is shown below, in the strongly polarized regime the three-point vertex function of impurity problem $\gamma'_{\nu\omega}$ that connects two fermionic propagators and interaction is close to unity~\eqref{eq:vertex_unity}. Then, the local polarization function of impurity can be approximated as $\Pi_{\omega} \simeq \chi^{0}_{\omega}$ and the full local susceptibility in the spin channel reads
\begin{align}
\chi^{-1}_{\omega} = \chi^{0~-1}_{\omega} - {\cal U}^{\pm}_{\omega}.
\end{align}
Here, ${\cal U}^{\pm}_{\omega}=-U+\Lambda$ is the bare interaction of impurity problem in the spin channel.
Then, the averaged 2PI four-point vertex function~\eqref{eq:gamma2PIapp} reads $\overline\gamma^{\rm 2PI}_{\av{\nu\nu'}\omega}\simeq{}U-\Lambda$ and one finally gets the following simple expression for the magnetic susceptibility~\eqref{eq:X_DMFTapp}
\begin{align}
X^{\rm ladd}_{\qv\omega} = X^{0}_{\qv\omega} \left[1+\left(U + J^{\rm d}_{\bf q}\right)X^{0}_{\qv\omega}\right]^{-1}.
\end{align}

\section*{Application: The Hubbard model on the hypercubic lattice in infinite dimensions}

We consider the half-filled Hubbard model 
\begin{align}
H=-(2d)^{-\frac{1}{2}}\sum_{\langle ij\rangle\sigma} c_{i\sigma}^{\dagger} c_{j\sigma}+U\sum_{i} n_{i\uparrow}n_{i\downarrow}
\label{eq:Hubbard}
\end{align}
on the hypercubic lattice in infinite dimensions $d\rightarrow\infty$, where the summation over $\av{ij}$ runs over nearest neighbors. In this limit the non-interacting density of states becomes a Gaussian, $D(\epsilon)=(2\pi)^{-1/2}e^{-\epsilon^2/2}$~\cite{RevModPhys.68.13}. At low temperatures this system favors antiferromagnetic order over paramagnetism. Within the symmetry-broken phase one has to consider two sublattices $A$ and $B$ of the bipartite hypercubic lattice with a staggered magnetization, $\langle m^{A}\rangle=-\langle m^{B}\rangle=\av{m}$. In a bipartite ordered state the volume of the Brillouin zone (BZ) is halved, such that Fourier transforms may only be performed up
to the magnetic unit cell, see, for example,~\cite{RevModPhys.68.13}. 
In the reduced Brillouin zone (RBZ) the noninteracting Hamiltonian reads,
\begin{align}
    H_0=\sum_{\kv\sigma}(a^{*}_{\kv\sigma},b^{*}_{\kv\sigma})
    \begin{pmatrix}
        0 & \varepsilon_\kv \\
          \varepsilon_\kv  & 0
    \end{pmatrix}
    \begin{pmatrix}
        a_{\kv\sigma}\\
        b_{\kv\sigma}
    \end{pmatrix},
    \label{app:hmlt}
\end{align}
where $a^{(*)}_{\kv\sigma}$ and $b^{(*)}_{\kv\sigma}$ annihilate (create) a $\sigma$-electron with momentum $\kv$ in sublattice A and B, respectively.
$\varepsilon_\kv$ is the dispersion of the hypercubic lattice and $\kv$ a vector of the RBZ. Therefore, the Green's function becomes a two-by-two matrix $\hat{G}=(G^{AA}, G^{AB}; G^{BA}, G^{BB})$ in sublattice space. Since the Hubbard model in infinite dimensions is an exact limit of DMFT, the lattice model~\eqref{eq:Hubbard} is mapped exactly to a single-site Anderson impurity model (AIM). Therefore, the self-energy $\Sigma_{\nu\sigma}$ of $\hat{G}_{\kv\nu\sigma}$ is local and it reads
\begin{align}
\hat{G}_{\kv\nu\sigma}=
\begin{pmatrix}
\zeta_{\nu\msigma} & -\varepsilon_\kv \\
-\varepsilon_\kv  & \zeta_{\nu\sigma}
\end{pmatrix}^{-1},
\end{align}
where $\zeta_{\nu\sigma}=i\nu+\mu-\Sigma_{\nu\sigma}$. The impurity $\hat{g}_{\nu\sigma}$ and local part of the lattice Green's function $\hat{G}_{\text{loc}}=\sum_\kv\hat{G}_\kv$ are tied via the following prescription
\begin{align}
&\hat{G}_{\text{loc},\nu\sigma}=\hat{g}_{\nu\sigma}=\int_{-\infty}^\infty \frac{D(\epsilon)d\epsilon}{\zeta_{\nu\sigma}\zeta_{\nu\msigma}-\epsilon^2}\begin{pmatrix}
\zeta_{\nu\msigma} & 0 \\
0  & \zeta_{\nu\sigma}
\end{pmatrix}.
\end{align}
The momentum summation was rewritten as an integral over the density of states $D(\epsilon)$ of the hypercubic lattice. It was used that the off-diagonal elements of Green's function are an odd function of $\epsilon$ and thus vanish upon integration. By symmetry, an exchange of the sublattice indices $A\leftrightarrow B$ is equivalent to a flip of the spin label $\sigma\leftrightarrow\msigma$. The prescription is satisfied by fixing the dynamical Weiss field $\mathcal{G}^{-1}_{\nu\sigma}=G^{-1}_{\text{loc},\nu\sigma}+\Sigma_{\nu\sigma}$ of the AIM self-consistently.

\subsection*{DMFT susceptibility of the ordered phase}

In order to calculate the transversal spin susceptibility of the Hubbard model in the antiferromagnetically ordered phase we introduce the bare susceptibility.
On a bipartite lattice it is in general necessary to consider two-particle quantities with four indices $a,b,c,d$. The bubble is then a $4\times4$ matrix given by the tensor product $\hat{G}_\up\otimes\hat{G}^T_\dn$. The locality of the irreducible vertex in DMFT allows to consider the Bethe-Salpeter equation only in a $2\times2$ subspace, where the bare susceptibility is given by the following point-wise product $\hat{G}_\up\circ\hat{G}^T_\dn$
\begin{align}
\hat{X}^{0}_{\qv\omega;\nu}=&\sum_\kv \hat{G}_{\kv+\qv,\nu+\omega\up}\circ\hat{G}^T_{\kv\nu\dn}\notag\\
=&\sum_\kv \frac{1}{\zeta_{\nu+\omega,\up}\zeta_{\nu+\omega,\dn}-\varepsilon^2_{\kv+\qv}} \frac{1}{\zeta_{\nu\dn}\zeta_{\nu\up}-\varepsilon^2_\kv}
\begin{pmatrix}
\zeta_{\nu+\omega,\dn}\zeta_{\nu,\up}   & \varepsilon_{\kv+\qv}\varepsilon_\kv \\
\varepsilon_{\kv+\qv}\varepsilon_\kv  & \zeta_{\nu+\omega,\up}\zeta_{\nu,\dn}
\end{pmatrix}\nonumber\\
=&\iint_{-\infty}^\infty d\epsilon_1d\epsilon_2
\frac{D_{\qv}(\epsilon_1,\epsilon_2)}{(\zeta_{\nu+\omega,\up}\zeta_{\nu+\omega,\dn}-\epsilon^2_1)(\zeta_{\nu\dn}\zeta_{\nu\up}-\epsilon^2_2)}
\begin{pmatrix}
\zeta_{\nu+\omega,\dn}\zeta_{\nu,\up}   & \epsilon_1\epsilon_2 \\
\epsilon_1\epsilon_2& \zeta_{\nu+\omega,\up}\zeta_{\nu,\dn}
\end{pmatrix},
\label{app:bubble}
\end{align}
where $\kv$ and $\qv$ are vectors of the reduced Brillouin zone (RBZ). Here, the momentum summation leads to a double integral involving the expression $D_{{\qv}}(\epsilon_1,\epsilon_2)$. This reduces to $D(\epsilon_1)D(\epsilon_1)$ for any generic wave vector $\qv$. The term ``generic'' may be understood such that $\qv$ is a vector of the RBZ with an infinite number of random entries (see~\cite{RevModPhys.68.13} and references therein).
As a consequence, the integrals in Eq.~\eqref{app:bubble} factorize and the bubble is given as $\hat{g}_{\nu+\omega\up}\circ\hat{g}_{\nu\dn}$.
Hence, the nonlocal bubble
\begin{align}
\tilde{X}^{0,ab}_{\qv\omega;\nu}=\sum_\kv G^{ab}_{\kv+\qv,\nu+\omega,\up}G^{ba}_{\kv\nu\dn}-g^a_{\nu+\omega,\up}g^a_{\nu\dn}\delta_{ab},
    \label{def:nonlocalbubble}
\end{align}
vanishes identically at generic $\qv$. In the following, we consider the non-generic vector $\qv_0=\mathbf{0}$ of the RBZ, where $D_{\qv_{0}}(\epsilon_1,\epsilon_2)=\delta(\epsilon_1-\epsilon_2)D(\epsilon_1)$. This may be used to eliminate one of the integrals in Eq.~\eqref{app:bubble}, the remaining integral is solved numerically. Vectors $\tilde{\qv}$ of the Brillouin zone (BZ) are marked with a tilde.

From the nonlocal bubble in Eq.~\eqref{def:nonlocalbubble} and from the full local four-point vertex of impurity model
$\overline{\gamma}^{ab}_{\nu\nu'\omega}=\overline{\gamma}^{\,a}_{\nu\nu'\omega}\, \delta_{ab}$ one obtains the
$T$-matrix $F^{ab}_{\qv\omega; \nu\nu'}$ via the Bethe-Salpeter equation (BSE),
\begin{align}
\hat{F}^{-1}_{\qv\omega} = \hat{\overline{\gamma}}^{-1}_{\omega} +\hat{\tilde{X}}^{0}_{\qv\omega},
\label{eq:bse}
\end{align}
where $\tilde{X}^{0,ab}_{\qv\omega; \nu\nu'}=\tilde{X}^{0,ab}_{\qv\omega;\nu}\,\delta_{\nu\nu'}$ and $\hat{O}^{-1}$ denotes a super-matrix inversion
with respect to the indices $(a,\nu)$ and $(b,\nu')$.
One further calculates the dual polarization in the ladder approximation~\eqref{eq:Piladdapp} as
\begin{align}
\hat{\tilde{\Pi}}^{\rm ladd}_{\qv\omega}=\text{Tr}_{\nu\nu'}
\left[\hat{\gamma}_{\omega}\hat{\tilde{X}}^{0}_{\qv\omega}\left(I - \text{V.C.}\right)\hat{\gamma}_{\omega}\right],
\label{eq:dualpolarization}
\end{align}
where $\text{Tr}_{\nu\nu'}$ denotes a trace over fermionic frequencies and 
V.C. indicates vertex corrections given by $\hat{F}_{\qv\omega}\hat{\tilde{X}}^{0}_{\qv\omega}$. Finally, the lattice susceptibility is obtained using the relation~\eqref{eq:XDMFTapp}. 
Further, we consider an approximation for the magnetic susceptibility given by equation~\eqref{eq:X1pmapp} in the case of Hubbard model
\begin{align}
\left[X^{(2)}_{\qv\omega}\right]^{-1}
&= \chi^{-1}_{\omega} - \tilde{\Pi}^{(2)}_{\qv\omega},
\label{eq:XDMFTapproximatedapp}
\end{align}
where $\tilde{\Pi}^{(2)}_{\qv\omega}$ is obtained when neglecting vertex corrections in Eq.~\eqref{eq:dualpolarization}. The case of magnetic susceptibility~\eqref{eq:XDMFTapp} where the polarization function $\tilde{\Pi}^{\rm ladd}_{\qv\omega}$ is approximated by the second-order correction $\tilde{\Pi}^{(2)}_{\qv\omega}$ is also considered. However, is does not provide a good approximation for the exact result of Eq.~\ref{eq:dualpolarization} as shown in the main text.

\subsection*{Numerical calculations}
\begin{figure}
\includegraphics[width=0.6\textwidth]{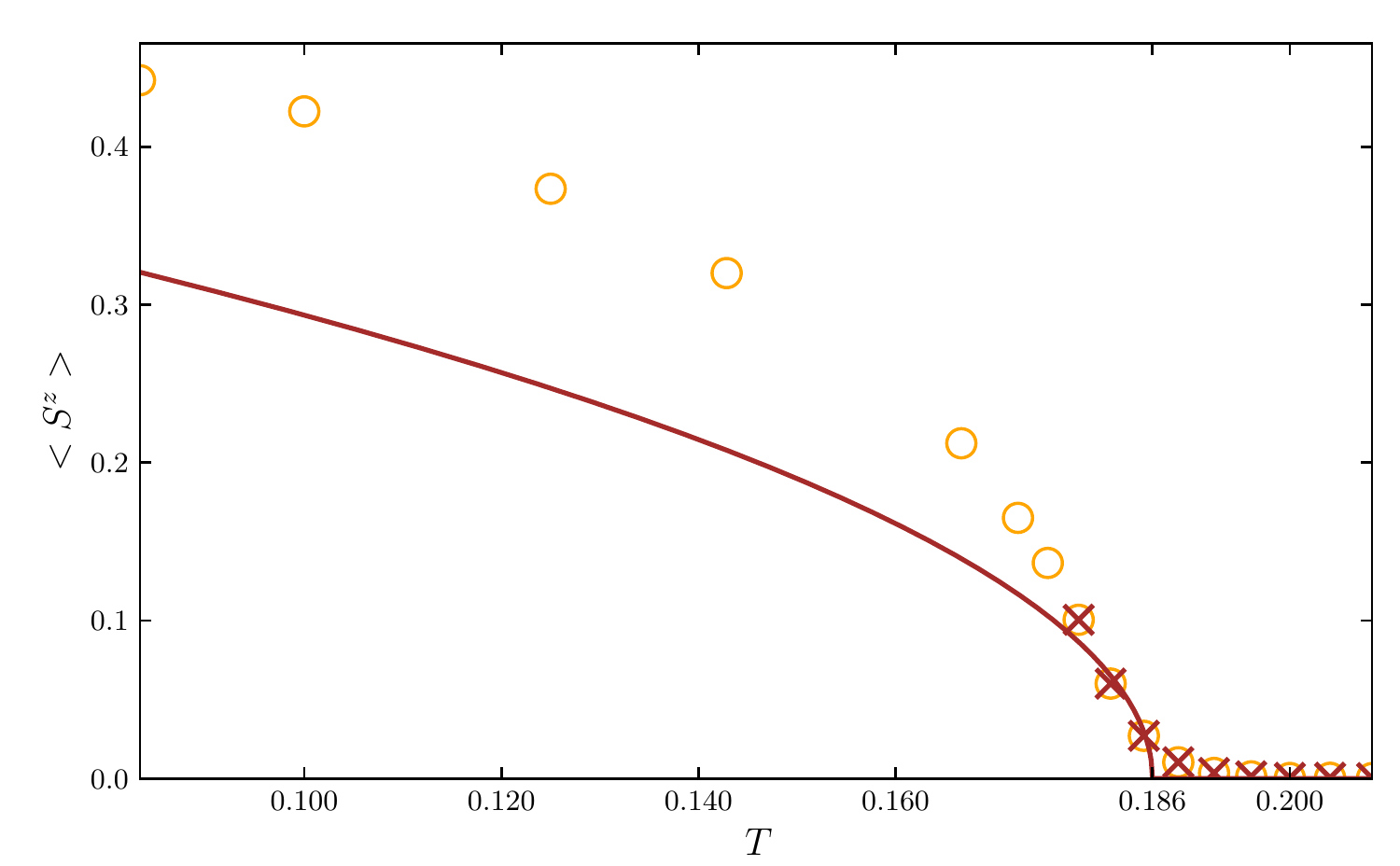}
\caption{The magnetic moment as a function of the temperature for the interaction value of $U=5$. The mean-field model(solid line) fits DMFT results close to the transition(red crosses). DMFT data for lower temperatures are shown, too(circles).}
\label{fig:tneelfit}
\end{figure}

The numerical calculations are performed using $~10^8$ measurements with $50$ Monte-Carlo moves between them. Aside from the segment insertion and removal we also use the shift and the double move as well as the spin-flip, a global move. We measure the Green's function in the Legendre basis with $35$ coefficients. The Hilbert-transform for the local Green's function is done on an energy mesh of $\omega \in [-20, 20]$ with $4000$ mesh points. The initial DMFT-cycle is performed with an external magnetic field, that is switched off for the following cycles. For the DMFT updates we use a mixing parameter of $0.5$.

We fit the model $\av{S^z} = \sqrt{T_N - T}$ to the DMFT results to estimate $T_N\approx 0.186$, see Fig.~\ref{fig:tneelfit}. This refined scan is done for $U=5$ with a more dense temperature mesh. Only data points in proximity to the transition were taken into account and the domain of the paramagnetic region is treated by a Heaviside step-function.

For simplicity, we calculate an approximated versions~\eqref{eq:X2approxapp} and~\eqref{eq:X_DMFTapp} of the magnetic susceptibilities~\eqref{eq:X1pmapp} and~\eqref{eq:X_DBapp} for the Hubbard model ($V_{\qv}, J^{\rm d}_{\qv}, \Lambda=0$), respectively. We note that the wave vector $\qv_0$ of the RBZ maps to two vectors $\tilde{\qv}_0=\qv_0$ and $\tilde{\qv}_\pi=(\pi,...,\pi)$ of the BZ. In the paramagnet this mapping diagonalizes the susceptibility matrix $\hat{X}(\qv_0)=(X^{AA}, X^{AB}; X^{BA}, X^{BB})$, where the diagonal elements are $X(\tilde{\qv}_\pi)=X^{AA}+X^{BB}-X^{AB}-X^{BA}$ and $X(\tilde{\qv}_0)=X^{AA}+X^{BB}+X^{AB}+X^{BA}$. In the ordered phase the same mapping does not diagonalize $\hat{X}$, since the offdiagonal element $X^{\pm}(\tilde{\qv}_0)=X^{AA}-X^{BB}+X^{AB}-X^{BA}$ does not vanish. Approaching $T_N$ from above, $X(\tilde{\qv}_\pi)$ diverges and it remains divergent in the ordered phase, signaling that the crystal is prone to a spontaneous tilt of its magnetization axis. We verified in our calculations that at $U=5$ and $T=0.1<T_N$ one of the two eigenvalues of the BSE~\eqref{eq:bse} is very close to unity, $|\lambda_{\qv_\pi}|\approx0.993$, and that this channel corresponds to $X(\tilde{\qv}_\pi)$. We account the slight deviation of this eigenvalue from unity to our approximation of the impurity vertex $\gamma$. The second eigenvalue, however, remains smaller than one, $|\lambda_{\qv_0}|\approx0.715$, and belongs to the homogenous susceptibility $X(\tilde{\qv}_0)$. In the main text we show the real and imaginary parts of the lattice susceptibility $\hat{X}(\tilde{\qv}_0)$, which corresponds to $\Re \hat{X}(\tilde{\qv}_0) = \Re X(\tilde{\qv}_0)$ and $\Im \hat{X}(\tilde{\qv}_0) = \Im X^{\pm}(\tilde{\qv}_0)$.

\subsection*{Ward identity}
We deduce two exact statements about the dynamical homogenous susceptibility $\hat{X}(\qv_0,\omega)$ from the Ward identity of the two-particle correlation function $G^{abcd}_{kk'q}=-\avsmall{a^{\phantom{*}}_{k\dn}b^{*}_{k+q,\up}c^{\phantom{*}}_{k'+q,\up}d^{*}_{k'\dn}}$, where each of the operators labeled with $a,b,c,d$ denotes either $a^{(*)}$ or $b^{(*)}$, respectively, and $k=(\kv,\nu), q=(\qv,\omega)$ abbreviate momenta from the RBZ and Matsubara frequencies. From the equation of motion $\partial_\tau\rho^a_\qv=[H,\rho^a_\qv]$ of the density operator
$\rho^{a}_\qv=\sum_\kv a^{*}_{\kv\dn}a^{\phantom{*}}_{\kv+\qv\up}$ one obtains the Ward identity (see, for example,~\cite{KrienF}),
\begin{align}
-i\omega\sum_{k'}\left(G^{aaaa}_{kk'q}+G^{aabb}_{kk'q}\right)
+\sum_{k'}[\varepsilon_{k'+q}-\varepsilon_{k'}]\left(G^{aaba}_{kk'q}+G^{aaab}_{kk'q}\right)
=G^{aa}_{k+q\up}-G^{aa}_{k\dn},\;\;(a\neq b),\label{eq:wardid}
\end{align}
where $\sum_k$ implies a summation over the RBZ and Matsubara frequencies.
Evaluating Eq.~\eqref{eq:wardid} at $q_0^+=(\qv_0=\mathbf{0},\omega^+>0)$ the term in the second line vanishes.
Upon summation over $k$ and using that $\sum_{kk'}G^{aabb}_{kk'q_0^+}=X^{ab}(\qv_0,\omega^+)$ it follows,
\begin{align}
-i\omega^+\left[X^{aa}(\qv_0,\omega^+)+X^{ab}(\qv_0,\omega^+)\right]=\av{m^a},
\end{align}
where it was also used that $\sum_k G^{aa}_{k\sigma}=\av{n^a_\sigma}$ and $\av{m^a}=\langle n^a_\up\rangle-\langle n^a_\dn\rangle$.
Adding up above relation for $a=A, b=B$ and $a=B, b=A$ it follows that
\begin{align}
    &X(\tilde{\qv}_0,\omega^+)=X^{AA}(\qv_0,\omega^+)+X^{AB}(\qv_0,\omega^+)+X^{BB}(\qv_0,\omega^+)+X^{BA}(\qv_0,\omega^+)=0,\label{eq:homexact}
\end{align}
since $\av{m^A}=-\av{m^B}=\av{m}$. Subtraction likewise leads to
\begin{align}
    &X^\pm(\tilde{\qv}_0,\omega^+)=X^{AA}(\qv_0,\omega^+)+X^{AB}(\qv_0,\omega^+)-X^{BB}(\qv_0,\omega^+)-X^{BA}(\qv_0,\omega^+)=\frac{2i\av{m}}{\omega^+}.\label{eq:pmexact}
\end{align}
Eqs.~\eqref{eq:homexact} and~\eqref{eq:pmexact} follow from the equation of motion of the total spin density, $\rho^A_{\qv_0}+\rho^B_{\qv_0}$,
and are therefore necessary criteria for global spin conservation.

\section*{Spin polarized solution of atomic problem}
One can perform an exact diagonlization of a magnetically polarized single orbital Hubbard atom at the half-filling. The thermodynamic potential operator of the atom is given by
\begin{equation}
\hat{H}-\mu\hat{N}=
\sum_{\sigma}\Delta^{\phantom{*}}_{\sigma}
c^{*}_{\sigma}c^{\phantom{*}}_{\sigma}
+Un_{\uparrow}n_{\downarrow},
\end{equation}
Here $\Delta_{\uparrow,\downarrow}=\pm B-\mu$ and the magnetic field $B$ is considered much larger than the temperature $T\equiv1/\beta$.    
The system has four eigenstates $\left|0\right>$, $\left|\uparrow\right\rangle$, $\left|\downarrow\right\rangle$ and $\left|\uparrow\downarrow\right\rangle$ with the corresponding energies $E_0=0$, $E_{\uparrow,\downarrow}=\pm B-\mu$ and $E_{\uparrow\downarrow}=U-2\mu$. Half-filling corresponds to $\mu=U/2$, so that $E_{\uparrow\downarrow}=0$. Indeed, the partition function for $\mu=U/2$ is
\begin{equation}
Z=2+e^{\beta(\mu-B)}+e^{\beta(\mu+B)}\approx e^{\beta(B+\mu)},
\end{equation}
and the average filling is given by $(1\times{}e^{\beta(\mu-B)}+e^{\beta(\mu+B)}+2\times{}1)/Z=1$, where we used that $B\beta\gg1$.

The non-zero matrix elements of the creation and annihilation operators are
\begin{align}
\left<\uparrow\right|c^{*}_{\uparrow}\left|0\right>=1;~~ 
\left<\downarrow\right|c^{*}_{\downarrow}\left|0\right>=1;~~
\left<\uparrow\downarrow\right|c^{*}_{\uparrow}\left|\downarrow\right>=1;~~
\left<\uparrow\downarrow\right|c^{*}_{\downarrow}\left|\uparrow\right>=-1.
\end{align}
    
Now we use the Lehmann representation to obtain the Green's functions of the system
\begin{align}
g_{\nu\sigma}=\frac{1}{Z}\sum_{ij} \left|\left<i\right|c_{\sigma}\left|j\right>\right|^2\frac{e^{-\beta{}E_i}+e^{-\beta{}E_j}}{i\nu+E_i-E_j}.
\end{align}
This yields
\begin{align}
g_{\nu\uparrow}&=\frac{1}{Z}\left[\left|\left<0\right|c_{\uparrow}\left|\uparrow\right>\right|^2\frac{1+e^{\beta(\mu-B)}}{i\nu+\mu-B}+\left|\left<\downarrow\right|c_{\uparrow}\left|\uparrow\downarrow\right>\right|^2\frac{1+e^{\beta(\mu+B)}}{i\nu-\mu-B}\right],\\
g_{\nu\downarrow}&=\frac{1}{Z}\left[\left|\left<0\right|c_{\downarrow}\left|\downarrow\right>\right|^2\frac{1+e^{\beta(\mu+B)}}{i\nu+\mu+B}+\left|\left<\uparrow\right|c_{\downarrow}\left|\uparrow\downarrow\right>\right|^2\frac{1+e^{\beta(\mu-B)}}{i\nu-\mu+B}\right], \notag
\end{align}
or, using $B\beta\gg1$,
\begin{equation}
g_{\nu\uparrow}=\frac{1}{i\nu-\mu-B}\simeq\frac{1}{i\nu-U/2},~~~~
g_{\nu\downarrow}=\frac{1}{i\nu+\mu+B}\simeq\frac{1}{i\nu+U/2}.
\label{eq:Gat}
\end{equation}
Now we calculate the magnetic susceptibility
\begin{align}
\chi^{-+}_{\omega}=-\frac{1}{Z} \int_0^{\beta}d\tau\,e^{i\omega\tau}\av{S^{-}(\tau)\,S^{+}(0)},
\end{align}
where $S^{\pm}(\tau)$ are Heisenberg representations of $S^{\pm}$ operators.
The non-zero matrix elements of the latter are
\begin{equation}
\av{\uparrow\left|S^+\right|\downarrow}=\av{\downarrow\left|S^-\right|\uparrow}=1.
\end{equation}
Lehmann representation reads
\begin{equation}
\chi^{-+}_{\omega}=-\frac{1}{Z}\sum_{ij} \left|\av{i\right|S^-\left|j}\right|^2\left[\frac{\left(e^{-\beta{}E_j}-e^{-\beta{}E_i}\right)(1-\delta_{E_i,E_j})}{i\omega+E_i-E_j}+\beta\delta_{E_i,E_j}\delta_{\omega0}\right] =\frac{1}{i\omega-2B}.
\end{equation}

Finally we turn to calculating of $\chi^{\downarrow\uparrow+}(\tau_1,\tau_2)\equiv\av{T_{\tau}c^{*}_{\downarrow}(\tau_1)
c^{\phantom{*}}_{\uparrow}(\tau_2)S^+(0)}$.
Unlike the previous cases, here we have to explicitly consider the time-ordering operator.
\begin{align}
\chi^{\sigma\sigma'\varsigma}(\tau_1,\tau_2)=\av{c^{*}_{\sigma}(\tau_1) c^{\phantom{*}}_{\sigma'}(\tau_2)\,S^{\varsigma}(0)}\theta(\tau_1-\tau_2) - \av{c^{\phantom{*}}_{\sigma'}(\tau_2)c^{*}_{\sigma}(\tau_1)\,S^{\varsigma}(0)}\theta(\tau_2-\tau_1).
\end{align}
The usual trick here is to split the integration region $0<\tau_1,\tau_2<\beta$ in the Fourier transform integral into two parts: $\int_0^{\beta}\int_0^{\beta}d\tau_1d\tau_2\dots=\int_0^{\beta}d\tau_1\int_0^{\tau_1}d\tau_2\dots+\int_0^{\beta}d\tau_2\int_0^{\tau_2}d\tau_1\dots$ and swapping the integration variables in the second term. This immediately gives for
\begin{align}
\chi^{\sigma\sigma'\varsigma}(\nu_1,\nu_2)\equiv\int_0^{\beta}
\int_0^{\beta}d\tau_1d\tau_2
e^{i\nu_1\tau_1+i\nu_2\tau_2}\chi^{\sigma\sigma'\varsigma}(\tau_1,\tau_2)
\end{align}
the following Lehmann representation
\begin{align}
\chi^{\sigma\sigma'\varsigma}(\nu_1,\nu_2)=\frac{1}{Z}\left[\sum_{ijk}\av{i\right|c^{*}_{\sigma}\left|j}\av{j\right|c^{\phantom{*}}_{\sigma'}\left|k}\av{k\right|\,S^{\varsigma}\left|i}f_{ijk}(\nu_1,\nu_2)
-\sum_{ijk}\av{i\right|c^{\phantom{*}}_{\sigma'}\left|j}\av{j\right|c^{*}_{\sigma}\left|k}\av{k\right|\,S^{\varsigma}\left|i}f_{ijk}(\nu_2,\nu_1)\right], \label{chigen}
\end{align}
where
\begin{align}
f_{ijk}(\nu_1,\nu_2)
&= \int_0^{\beta} d\tau_1 \int_0^{\tau_1} d\tau_2 e^{-\beta{}E_i} e^{i\nu_1\tau_1+i\nu_2\tau_2} e^{\tau_1(E_i-E_j)+\tau_2(E_j-E_k)} \\
&=\frac{(e^{-\beta{}E_k}-e^{-\beta{}E_i})(1-\delta_{E_i,E_k})}{(i\nu_1+i\nu_2+E_i-E_k)(i\nu_2+E_j-E_k)}+\frac{e^{-\beta{}E_j}+e^{-\beta{}E_i}}{(i\nu_1+E_i-E_j)(i\nu_2+E_j-E_k)}
+\frac{\beta\delta_{E_i,E_k}\delta_{\nu_1+\nu_2,0}}{i\nu_2+E_j-E_k}.
\label{fgen}
\end{align}
For our particular case $\sigma=\downarrow$, $\sigma'=\uparrow$ and $\varsigma=+$, so $i=\left|\downarrow\right>$, $k=\left|\uparrow\right>$ and $j$ can be either $\left|0\right>$ for the first term in (\ref{chigen}) or $\left|\uparrow\downarrow\right>$ for the second one. Thus
\begin{equation}
\chi^{\downarrow\uparrow+}(\nu_1,\nu_2)=\frac{1}{Z}\left(f_{\downarrow,0,\uparrow}(\nu_1,\nu_2)+f_{\downarrow,\uparrow\downarrow,\uparrow}(\nu_2,\nu_1)\right).
\end{equation}
Using (\ref{fgen}) and $\beta B\gg1$ we obtain
\begin{align}
\chi^{\downarrow\uparrow+}(\nu_1,\nu_2)&=\left(1-\frac{2\mu}{i\nu_1+i\nu_2-2B}\right)\frac{1}{(i\nu_1-B-\mu)(i\nu_2-B-\mu)}\\
&=-g_{\nu_1\downarrow}g_{-\nu_2\uparrow}\left(1-U\chi^{-+}(\nu_1+\nu_2)\right).
\end{align}
Let us define the three-point vertex $\gamma'_{\nu\omega}$ for the spin channel that connects two fermionic propagators and interaction in the same way as in~\cite{PhysRevB.94.205110} with the cut-off on the renormalization parameter $\alpha^{-+}_{\omega} = \left(1+U^{-+}\chi^{-+}(\nu_1+\nu_2)\right)={\cal W}^{-+}_{\omega}/U^{-+}$ instead of $\chi^{-+}_{\omega}$. The difference between these two definitions is that in the case of $\gamma'_{\nu\omega}$ the full bosonic propagator of the impurity problem that is attached to the vertex is the full local susceptibility $\chi_{\omega}$, while in the case of $\gamma'_{\nu\omega}$ vertex function it is equal to the renormalized interaction of impurity problem ${\cal W}^{-+}_{\omega}$ in the spin channel. 

Remarkably, the three-point vertex function $\gamma'_{\nu\omega}$ in the spin-polarized case is equal to unity
\begin{align}
\gamma'^{\,-}(\nu_1,\nu_2)= \frac{-\av{c^{*}_{\downarrow}(\nu_1)\,c^{\phantom{*}}_{\uparrow}(-\nu_2)\,S^{\,+}(\nu_1+\nu_2)}}
{g_{\nu_1\downarrow}g_{-\nu_2\uparrow}\,\alpha^{-+}(\nu_1+\nu_2)}=1,
\label{eq:vertex_unity}
\end{align}
because in the spin channel the bare interaction is equal to $U^{+-}=-U$.
Using the relation between the three- and four-point vertices derived in~\cite{PhysRevB.94.205110}, one gets
\begin{align}
\gamma'^{\,\varsigma}_{\nu\omega} &=
\alpha^{\,\varsigma~-1}_{\omega} \sum_{\nu'}
\left[1 - \overline{\gamma}^{\,\varsigma}_{\nu\nu'\omega}g_{\nu'\sigma} g_{\nu'+\omega,\sigma'}\right] \\
\gamma'^{\,\varsigma}_{\nu\omega}
\left(1+{\cal W}^{\,\varsigma}_{\omega}\,
\Pi^{\,\varsigma}_{\omega}\right) &=
\sum_{\nu'}\left[1 - \overline{\gamma}^{\,\varsigma}_{\nu\nu'\omega}\,g_{\nu'\sigma} g_{\nu'+\omega,\sigma'}\right] \\
\gamma'^{\,\varsigma}_{\nu\omega} &=
\sum_{\nu'}\left[1 - \left(\overline{\gamma}^{\,\varsigma}_{\nu\nu'\omega} +\gamma'^{\,\varsigma}_{\nu\omega}{\cal W}^{\,\varsigma}_{\omega}\gamma'^{\,\varsigma}_{\nu'+\omega,-\omega}\right)g_{\nu'\sigma} g_{\nu'+\omega,\sigma'}\right],
\end{align}
where the Hedin expression for the polarization function of impurity problem $\Pi_{\omega} = \sum_{\nu}\gamma'_{\nu+\omega,-\omega}\,g_{\nu,\sigma}g_{\nu+\omega,\sigma'}$ is used. Therefore, when the three-point vertex function $\gamma'_{\nu\omega}$ is close to unity, the main contribution to the four-point vertex function is given by the following expression 
\begin{align}
\overline{\gamma}^{\,\varsigma}_{\nu\nu'\omega}\simeq-\gamma'^{\,\varsigma}_{\nu\omega}{\cal W}^{\,\varsigma}_{\omega}\gamma'^{\,\varsigma}_{\nu'+\omega,-\omega}.
\end{align}
Transforming back to the definition of the three-point vertex function used in this Letter $\gamma'_{\nu\omega}\to\gamma_{\nu\omega}$, one also has to replace the full local bosonic propagator as ${\cal W}_{\omega}\to\chi_{\omega}$. Then, the final expression for the four-point vertex reads
\begin{align}
\overline{\gamma}^{\,\varsigma}_{\nu\nu'\omega}\simeq-\gamma^{\,\varsigma}_{\nu\omega}\,\chi^{\,\varsigma}_{\omega}\,\gamma^{\,\varsigma}_{\nu'+\omega,-\omega}.
\end{align}

\section*{Application: Exchange interaction in the strongly localized regime}

Here, we calculate the exchange interaction for the Hubbard model in the strongly localized regime $t\ll{}U$. For this reason, let us find the nonlocal Green's function as the first order correction to the atomic limit solution with respect to the hopping amplitude. Then, using the definition of the Green's function, one gets
\begin{align}
G_{ij\sigma} = \frac{1}{Z} \int D[c^{*},c]~ c^{*}_{i\sigma} c^{\phantom{*}}_{j\sigma}~ e^{-\sum_{i}S^{i}_{\rm at} - \sum_{i'j'\sigma'}t_{i'j'}c^{*}_{i'\sigma'} c^{\phantom{*}}_{j'\sigma'}} = 
\frac{1}{Z} \int D[c^{*},c]~ c^{*}_{i\sigma} c^{\phantom{*}}_{j\sigma}~ e^{-\sum_{i}S^{i}_{\rm at}} - 
\frac{1}{Z} \int D[c^{*},c]~ \sum_{i'j'}t_{i'j'}c^{*}_{i\sigma} c^{\phantom{*}}_{j\sigma}c^{*}_{i'\sigma'} c^{\phantom{*}}_{j'\sigma'}~ e^{-\sum_{i}S^{i}_{\rm at}}.
\end{align} 
Since the atomic action is purely local, the contribution to the nonlocal Green's function for $i\neq{}j$ is given only by the second term when $i=j'$ and $j=i'$, so $\sigma=\sigma'$. Then, the nonlocal Green's function can be rewritten as   
\begin{align}
G_{ij\sigma} = t_{ji}~ \frac{1}{Z^{i}_{\rm at}} \int D[c^{*},c]~ c^{*}_{i\sigma} c^{\phantom{*}}_{i\sigma}~ e^{-S^{i}_{\rm at}} \times \frac{1}{Z^{j}_{\rm at}} \int D[c^{*},c]~ c^{*}_{j\sigma} c^{\phantom{*}}_{j\sigma}~ e^{-S^{j}_{\rm at}} = t\,g^2_{\nu\sigma},
\end{align}    
where $g_{\nu\sigma}$ is the local Green's function of atomic problem. Taking into account the result of Eq.~\ref{eq:Gat} and that
the difference of the self-energies is equal to
$\Sigma_{\nu\uparrow} - \Sigma_{\nu\downarrow} = U+2B$ and that $2\av{S^{z}}=1$, the exchange interaction reads
\begin{align}
J_{ij} &= -\sum_{\nu} \left(\chi^{-1}_{\omega=0} + \frac{\Sigma_{\nu\uparrow}-\Sigma_{\nu\downarrow}}{2\av{S^{z}}}\right)
G^{\phantom{+}}_{ij,\nu\uparrow}G^{\phantom{+}}_{ji,\nu\downarrow} \left(\chi^{-1}_{\omega=0} + \frac{\Sigma_{\nu\uparrow}-\Sigma_{\nu\downarrow}}{2\av{S^{z}}}\right)
= -\sum_{\nu}\frac{tU}{(i\nu-U/2)^2}\frac{tU}{(i\nu+U/2)^2} 
=  -\frac{2t^2}{U}.
\end{align} 

\section*{Application: Antiferromagnetic dimer}
One can also perform an exact diagonlization of a two-site model with the antiferromagnetic ground state 
\begin{equation}
\hat{H}-\mu\hat{N}= t c^{*}_{1\sigma}c^{\phantom{*}}_{2\sigma} + t c^{*}_{2\sigma}c^{\phantom{*}}_{1\sigma} + 
\sum_{i=1,2;\,\sigma}\left(\Delta^{\phantom{*}}_{i\sigma}
c^{*}_{i\sigma}c^{\phantom{*}}_{i\sigma}
+Un_{i\uparrow}n_{i\downarrow}\right),
\end{equation}
where $\Delta_{\uparrow,\downarrow}=\pm B-\mu$ and the magnetic field $B$ is again considered much larger than the temperature $T\equiv1/\beta$. 
Using the Lehmann representation, one can obtain the nonlocal Green's functions at the low temperatures, i.e. $\beta{}U\gg1$ in the strongly-correlated regime $t\ll{}U$ as
\begin{align}
G_{ab,\sigma} = \frac{1}{Z}\sum_{ij}<\Psi_{i}\,|c^{*}_{a\sigma}|\,\Psi_{j}><\Psi_{j}\,|c^{\phantom{*}}_{b\sigma}|\,\Psi_{i}>\frac{e^{-\beta{E}_{i}}+e^{-\beta{E}_{j}}}{i\nu+E_{i}-E_{j}}
\end{align} 
Since only the low-lying energy states contribute to the Green's function at low temperatures, because the contribution of higher energy states is exponentially suppressed, we give only relevant energies and (unnormalized) eigenstates below\\
\begin{center}
\begin{tabular}{| l | l |}
\hline
~$E_{5} = -U/2+\sqrt{B^2+t^2}$ ~& 
~$\Psi_{5} = -\,\frac{-B - \sqrt{B^2 + t^2}}{t}\,|\uparrow\downarrow.\uparrow\,> +\,|\uparrow.\uparrow\downarrow\,>$ \\
\hline
~$E_{4} = -U/2+\sqrt{B^2+t^2}$ ~&
~$\Psi_{4} = -\,\frac{-B - \sqrt{B^2 + t^2}}{t}\,|\downarrow.~0> +\,|~0~.\downarrow\,>$ ~\\
\hline
~$E_{3\phantom{1}} = -U/2-\sqrt{B^2+t^2}$ ~& 
~$\Psi_{3\phantom{1}} = -\,\frac{-B + \sqrt{B^2 + t^2}}{t}\,|\uparrow\downarrow.\uparrow\,> +\,|\uparrow.\uparrow\downarrow\,>$ ~\\
\hline
~$E_{2\phantom{1}} = -U/2-\sqrt{B^2+t^2}$ ~& 
~$\Psi_{2\phantom{1}} = -\,\frac{-B + \sqrt{B^2 + t^2}}{t}\,|\downarrow.~0> +\,|~0~.\downarrow\,>$ ~\\
\hline
~$E_{1\phantom{1}} = -U-2B$ ~& 
~$\Psi_{1\phantom{1}} = |\uparrow\downarrow.~0> -\,\frac{2t}{U+2B}\,|\downarrow.\uparrow\,> -\,\frac{U+2B}{t} \,|\uparrow.\downarrow\,> +\,|~0~.\uparrow\downarrow\,>$~ \\
\hline
\end{tabular}
\end{center}
Then, one gets
\begin{align}
G_{ab\uparrow} &= \frac{1}{Z}<\Psi_{3}\,|c^{*}_{a\uparrow}|\,\Psi_{1}><\Psi_{1}\,|c^{\phantom{*}}_{b\uparrow}|\,\Psi_{3}> \frac{e^{-\beta{E}_{3}}+e^{-\beta{E}_{1}}}{i\nu+E_{3}-E_{1}} + \frac{1}{Z}<\Psi_{1}\,|c^{*}_{a\uparrow}|\,\Psi_{2}><\Psi_{2}\,|c^{\phantom{*}}_{b\uparrow}|\,\Psi_{1}> \frac{e^{-\beta{E}_{1}}+e^{-\beta{E}_{2}}}{i\nu+E_{1}-E_{2}} \\
&\,+ \frac{1}{Z}<\Psi_{5}\,|c^{*}_{a\uparrow}|\,\Psi_{1}><\Psi_{1}\,|c^{\phantom{*}}_{b\uparrow}|\,\Psi_{5}> \frac{e^{-\beta{E}_{5}}+e^{-\beta{E}_{1}}}{i\nu+E_{5}-E_{1}} + \frac{1}{Z}<\Psi_{1}\,|c^{*}_{a\uparrow}|\,\Psi_{4}><\Psi_{4}\,|c^{\phantom{*}}_{b\uparrow}|\,\Psi_{1}> \frac{e^{-\beta{E}_{1}}+e^{-\beta{E}_{4}}}{i\nu+E_{1}-E_{4}} \\
&= \frac{1}{Z}\frac{1}{N^2_{1}N^2_{3(5)}}
\left(-\,\frac{-B \pm \sqrt{B^2 + t^2}}{t}\,<\,\uparrow\downarrow.\uparrow| +\,<\,\uparrow.\uparrow\downarrow|\right)\left(-\,\frac{2t}{U+2B}\,|\uparrow\downarrow.\uparrow\,> +\,|\uparrow.\uparrow\downarrow\,>\right)\times \notag\\
&\times\left(-\,<\uparrow\downarrow.\uparrow| +\,\frac{U+2B}{t}\,<\uparrow.\uparrow\downarrow|\right)\left(-\,\frac{-B \pm \sqrt{B^2 + t^2}}{t}\,|\uparrow\downarrow.\uparrow\,> +\,|\uparrow.\uparrow\downarrow\,>\right)
\frac{e^{\beta(U/2\pm\sqrt{B^2+t^2})}+e^{\beta(U+2B)}}{i\nu-(U/2\pm\sqrt{B^2+t^2})+(U+2B)} \notag\\
&\,+ \frac{1}{Z}\frac{1}{N^2_{1}N^2_{2(4)}}
\left(<\,\downarrow.~0~| -\,\frac{U+2B}{t} \,<0~.\downarrow|\right)\left(-\,\frac{-B \pm \sqrt{B^2 + t^2}}{t}\,|\downarrow.~0> +\,|~0~.\downarrow\,>\right)\times \notag\\
&\times\left(-\,\frac{-B \pm \sqrt{B^2 + t^2}}{t}\,<\downarrow.~0~|\, +<0~.\downarrow|\right) 
\left(-\,\frac{2t}{U+2B}\,|\downarrow.~0> +\,|~0~.\downarrow\,>\right)
\frac{e^{\beta(U/2\pm\sqrt{B^2+t^2})}+e^{\beta(U+2B)}}{i\nu+(U/2\pm\sqrt{B^2+t^2})-(U+2B)}. \notag
\end{align} 
Note that all eigenstates $\Psi_{i}$ were normalized as $\frac{1}{N_{i}}\Psi_{i}$. Simplifying the previous equation one gets
\begin{align}
G_{ab\uparrow} &= \frac{1}{Z}\frac{1}{N^2_{1}N^2_{3(5)}}
\left(\frac{-2B \pm 2\sqrt{B^2 + t^2}}{U+2B}+1\right)\left(\frac{-B \pm \sqrt{B^2 + t^2}}{t} +\frac{U+2B}{t}\right)
\frac{e^{\beta(U/2\pm\sqrt{B^2+t^2})}+e^{\beta(U+2B)}}{i\nu+U/2+2B\mp\sqrt{B^2+t^2}} \notag\\
&\,+ \frac{1}{Z}\frac{1}{N^2_{1}N^2_{2(4)}}
\left(-\,\frac{-B \pm \sqrt{B^2 + t^2}}{t}-\,\frac{U+2B}{t}\right)\left(\frac{-B \pm \sqrt{B^2 + t^2}}{t}\frac{2t}{U+2B} + 1\right)
\frac{e^{\beta(U/2\pm\sqrt{B^2+t^2})}+e^{\beta(U+2B)}}{i\nu-U/2-2B\pm\sqrt{B^2+t^2}}.
\end{align} 
Taking into account that
\begin{align}
Z &= 4\left(1+e^{\beta{}U/2}\cosh\beta{}B\right)\simeq{}e^{\beta(U+2B)} \\
N^2_{1} &= 2 + \left(\frac{2t}{U+2B}\right)^2 + \left(\frac{U+2B}{t}\right)^2 \simeq \left(\frac{U+2B}{t}\right)^2 \\
N^2_{2(4)} &= N^2_{3(5)} = 1 + \left(\frac{-B\pm\sqrt{B^2+t^2}}{t}\right)^2 
\end{align}
one can finally get
\begin{align}
G_{ab\uparrow} &= \frac{t^2}{(U+2B)^2}\frac{1}{1 + \left(\frac{-B\pm\sqrt{B^2+t^2}}{t}\right)^2}
\frac{U \pm 2\sqrt{B^2 + t^2}}{U+2B}\frac{U+B \pm \sqrt{B^2 + t^2}}{t}
\frac{1}{i\nu+U/2+2B\mp\sqrt{B^2+t^2}} \notag\\
&\,- \frac{t^2}{(U+2B)^2}\frac{1}{1 + \left(\frac{-B\pm\sqrt{B^2+t^2}}{t}\right)^2}
\frac{U+B \pm \sqrt{B^2 + t^2}}{t}\frac{U \pm 2\sqrt{B^2 + t^2}}{U+2B}
\frac{1}{i\nu-U/2-2B\pm\sqrt{B^2+t^2}}.
\end{align} 
If $U\gg{}B$ and $U\gg{t}$, we get
\begin{align}
G_{ab\uparrow} &= \sum_{\pm}\frac{t}{U}\frac{1}{1 + \left(\frac{-B\pm\sqrt{B^2+t^2}}{t}\right)^2}\left[
\frac{1}{i\nu+U/2} - \frac{1}{i\nu-U/2}\right] = \frac{t}{U}\left[
\frac{1}{i\nu+U/2} - \frac{1}{i\nu-U/2}\right] .
\end{align} 
The same result can be found for 
\begin{align}
G_{ba\downarrow} &= \frac{t}{U}\left[
\frac{1}{i\nu+U/2} - \frac{1}{i\nu-U/2}\right] .
\end{align} 
Therefore, the exchange interaction reads
\begin{align}
J_{ab}= 
-\frac{U^2}{\beta}\sum_{\nu} G_{ab\uparrow}G_{ba\downarrow} = \frac{t^2}{2\pi}\int_{-\infty}^{+\infty}\frac{2\,dx}{(x-iU/2)(x+iU/2)} = - \frac{2t^2}{U}.
\end{align}

\end{document}